\documentclass[preprint,11pt,1p,authoryear]{elsarticle}

\usepackage[T1]{fontenc}
\usepackage[latin9]{inputenc}
\usepackage{amsmath}
\usepackage{amssymb}
\usepackage{amsfonts}
\usepackage{graphicx}
\usepackage{lineno}
\usepackage{caption}
\usepackage[activate={true,nocompatibility},final,tracking=true,kerning=true,spacing=true,factor=1000,stretch=20,shrink=20]{microtype}
\SetExtraSpacing{font=*}{.={500,500,-100}}
\usepackage{multirow}
\usepackage{booktabs}
\newcommand{\minitab}[2][l]{\begin{tabular}{#1}#2\end{tabular}}
\makeatletter
\let\start@align@nopar\start@align
\let\start@gather@nopar\start@gather
\let\start@multline@nopar\start@multline
\long\def\start@align{\par\start@align@nopar}
\long\def\start@gather{\par\start@gather@nopar}
\long\def\start@multline{\par\start@multline@nopar}
\makeatother
\begin{document}
\begin{frontmatter}
\title{\Large\bfseries Maximum disorder model
       for dense steady-state flow\\
       of granular materials}
%      for anisotropy
%      in dense granular materials during slow steady flow}
\author[kuhn]{Matthew R. Kuhn}%
%\thanks
\address[kuhn]{Donald P. Shiley School of Engineering,
                University of Portland,
                5000 N. Willamette Blvd.,
                Portland, Oregon, USA  97203,
                kuhn@up.edu, (503)943-7361}
%\maketitle
%
\begin{abstract}
A flow model is developed for dense shear-driven granular flow.
As described in the geomechanics literature, a critical state
condition is reached after sufficient shearing beyond
an initial static packing.
During further shearing at the critical state, the
stress, fabric, and density remain nearly constant,
even as particles are being continually rearranged.
The paper proposes a predictive framework for critical
state flow, viewing it as a condition of maximum
disorder at the micro-scale.
The flow model is constructed in a two-dimensional setting from
the probability density of the motions, forces, and orientations
of inter-particle contacts. %, which
%constitute a high-dimensional phase space
%of the contact forces, contact movements, and
%contact orientations.
%During shear-driven frictional flow of a dense granular material,
%a critical state condition is reached,
%in which the stress, fabric, and density are constant, 
%even as the particles are continually rearranged.
%The paper considers this conditions as being maximally disordered:
%particle arrangements, movements, and interactions are mixed and, 
%in a statistical sense, 
%unbiased and optimally dispersed.
%The model is developed within a two-dimensional setting of disks.
Constraints are applied to this probability density:
%in accordance with steady granular
%flow:
constant mean stress, constant volume,
consistency of the contact dissipation rate with
the stress work, and 
the fraction of sliding contacts.
The differential form of Shannon entropy, a measure of disorder, is
applied to the density,
%in the probability
%density of the contact forces, contact movements, and
%contact orientations.
%Shearing deformation is assumed to result entirely from
%the tangential motions of the particles, and particle 
%force-movement interactions
%are modeled as rigid-friction and unilateral.
%(4) other constraining information
%extracted from discrete element (DEM) simulations.
and the Jaynes formalism is used to find the
density of maximum disorder in the underlying phase space.
The resulting distributions of contact force, movement,
and orientation are compared with two-dimensional DEM
simulations of biaxial compression.
The model favorably predicts anisotropies of the contact orientations,
contact forces, contact movements, and the orientations of those contacts
undergoing slip.
The model also predicts the relationships between contact force magnitude
and contact motion.
The model is an alternative to affine-field descriptions of
granular flow.
%The predictions are less favorable for the probability density of
%contact force, for the fraction of sliding contacts, and
%for the magnitudes of contact motions.
\end{abstract}
\end{frontmatter}
%
%\linenumbers
%\setlength\linenumbersep{7mm}
%
\section{\large Introduction}
The critical state concept in geomechanics holds that 
dense granular materials,
when loaded beyond an initial static packing,
eventually attain a steady state condition of constant density,
fabric, and stress
%after sufficient shearing beyond the initial state
\citep{Schofield:1968a}.
This condition is often associated with
shear-driven flow and failure: granular avalanches,
landslides, tectonic faults, and failures of foundation
systems and embankments. 
As such, the critical state has received intense interest from
geologists, engineers, and physicists, who have devoted great
effort in understanding the state's underlying mechanics.
Density, fabric, and deviatoric stress 
at the critical state are known to depend upon the particles'
shapes and contact properties as well as on the mean stress
and intermediate principal stress \citep{Zhao:2013a}. 
Even so, the eventual bulk characteristics for
a given assembly are insensitive to the initial particle arrangement
and to the stress path that ends in the critical state: 
for example, materials that are initially either loose or dense
eventually arrive at the same density condition after sufficient shearing.
%\citep{Casagrande:1936a,Zhao:2013b}.
%\citep{Zhao:2013b}.
This convergent characteristic resembles that of thermal systems that
approach an equilibrium condition with 
sufficient passage of time.
\par
Another pervasive feature of critical state flow is
the continual and intense activity of grains at the micro-scale,
%with rapid alterations of their positions, of their
%contact network, and of the forces and displacements
%at their contacts.
yet this local tumult produces a monotony
in the bulk fabric, stress, and density.
Micro-scale activity
occurs in three ways: (a)~statically, as alterations of the
inter-particle contact forces, (b)~geometrically, as changes
in the particles' configuration and local density,
and (c)~topologically, as changes in the load-bearing contact network
among the particles.
In these three respects,
we view the critical state as the condition of maximum disorder
that emerges during sustained shearing.
Early work by \citet{Brown:2000b} investigated disorder
in the local density,
and a recent paper by the author explored topological disorder
at the critical state \citep{Kuhn:2014a}.
The current paper addresses statical disorder 
as expressed in a two-dimensional (2D) setting,
focusing on anisotropies and distributions
of the contacts' orientations, forces,
and movements. 
The analysis applies to the critical state flow of dry unbonded
frictional materials of sufficient density
to develop a load-bearing (persistently jammed) network of contacts
during slow (quasi-static, non-collisional) shearing.
Particles are assumed durable (non-breaking) and nearly rigid, such that
deformations of the particles are small, even in
the vicinity of their contacts.
\par
During flow,
granular materials have an internal % structure that is expressed in an
organization of movement and force, % over a range of scales,
an organization that is in some respects pronounced but in others subtle.
We briefly review these characteristics of 
%micro-scale organization in dense granular flow at
the critical state,
as observed in laboratory experiments, numerical simulations, or both.
\begin{enumerate}[{A.}1.]
\item
The motions of individual particles do not conform to an affine, mean
deformation field, and fluctuations from the mean field are
large and seemingly erratic
%\citep{Calvetti:1997a,Radjai:2002a,Kuhn:2003d,Utter2008a}.
\citep{Kuhn:2003d}.
The rates at which contacting particles approach or withdraw from each
other (i.e. contact movements in their normal directions)
are generally much smaller than those 
corresponding to an affine field. %, and the mean normal rate is
% very nearly zero.
In contrast,
the transverse, tangential movements between
contacting pairs are much larger than those
of affine deformation
\citep{Kuhn:2003d}.
As a result, bulk deformation is almost entirely attributed
to the tangential movements
of particles \citep{Kuhn:2004k}.
%of particles \citep{Kuhn:2004k,Kruyt:2007a}.
%
\item\label{item:ContactElasticity}
Strength, expressed as a ratio of the principal
stresses, is insensitive to the contacts' elastic
stiffness and to the mean stress, 
such that simulations of either soft or hard particles
%as well as simulations with different contact stiffnesses
exhibit similar strengths at the critical 
state \citep{Hartl:2008a,Kruyt:2014a}.
Fabric measures at the critical state
(fraction of sliding contacts, contact anisotropy, etc.) are also
insensitive to contact stiffness \citep{daCruz:2005a}.
\item
Particle rotations are large when compared with the bulk deformation rate
%\citep{Calvetti:1997a,Kuhn:2004k}.
\citep{Kuhn:2004k}.
%\citep{Calvetti:1997a,Dedecker:2000a,Kuhn:2004k}.
In particular, the rolling motions between
particles are much larger than the
%sliding movements \citep{Bardet:1994a,Kuhn:2003h}.
sliding movements \citep{Kuhn:2003h}.
\item
The contact network is relatively sparse, in that the
number of contacts within the load-bearing contact network is
sufficient to produce a static indeterminacy (hyperstaticity)
but with only a modest excess of contacts
%\citep{Thornton:1998a,Thornton:2000a,Majmudar2007a}.
\citep{Thornton:2000a}.
The excess very nearly corresponds to the number of contacts that
%are sliding \citep{Cundall:1983a,Kruyt:2007a}.
are sliding \citep{Kruyt:2007a}.
This modest indeterminacy
is consistent with observations of intermittent, sudden reductions of
stress, which result from periodic collapse events
that are occasioned by fresh slip events or loss of contacts
\citep{Pena:2008a}.
This condition of marginal hyperstaticity is referred to
as ``jammed'' within the granular physics community.
\end{enumerate}
\begin{enumerate}[{B.}1.]
\item
During critical state flow, the contact fabric is anisotropic,
with the normals of the contacts oriented predominantly in the
direction of the 
major principal compressive stress 
%\citep{Oda:1985a,Thornton:1986a,Rothenburg:1989a,Calvetti:1997a}.
\citep{Rothenburg:1989a}.
\item
The normal contact forces
are larger among those contacts oriented in the direction of
the major principal compressive stress; whereas,
the averaged tangential forces are larger for contact surfaces oblique
to the principal stress directions
%\citep{Rothenburg:1989a,Ouadfel:1999a,Kruyt:2003b,Majmudar:2005a}.
\citep{Rothenburg:1989a,Majmudar:2005a}.
\item
Deviatoric stress is primarily borne by the normal contact forces 
between particles; whereas, the tangential contact forces 
make a much smaller contribution to the deviatoric stress 
%\citep{Cundall:1983a,Thornton:1986a,Rothenburg:1989a,Thornton:2000a,Kuhn:2003h}.
\citep{Thornton:2000a}.
\item
When considering only the normal forces, deviatoric stress 
is primarily carried by those contacts with
forces that are larger than the mean force (strong contacts), 
whereas the
remaining (weak) contacts contribute far less to the deviatoric 
stress \citep{Radjai:1998a,Kruyt:2007a}.
\item
Anisotropy of the contact network is also largely attributed to strong
contacts, which are predominantly oriented in the direction of
the major principal stress \citep{Radjai:1998a}.
\item
Many contacts slide in the ``wrong direction'' with respect
to the direction 
that corresponds to an affine deformation \citep{Kuhn:2003d}.
\item
Compared with other orientations,
contacts that are oriented in the direction of extension have a
greater average \emph{magnitude} of slip, but the mean slip velocity
is largest among contacts that are oriented obliquely to
the directions of compression and extension
\citep{Kuhn:2004k}.
\item
Frictional sliding is more common among those contacts with a 
smaller-than-mean normal force \citep{Radjai:1998a}.
\item
The more mobile contacts --- those
with large sliding movements --- tend to be those that bear 
a smaller-than-average normal force (i.e., weak contacts)
\citep{Kruyt:2007a}.
\item
Deformation, when measured at the meso-scale of particle clusters,
is related to contact orientation:
contacts with branch vectors that are more aligned with
that of bulk compression tend to produce local dilation;
whereas, contacts that are more aligned with the direction
of bulk extension tend to produce local compression.
These trends have been determined by studying the elongations of
voids that are surrounded by rings of particles and their 
branch vectors \citep{Nguyen:2009a}.
%
%\item
%Just as deviator stress results primarily from the more
%heavily loaded contacts (item~B.4),
%deformation is primarily due to the more active, ``mobile'' contacts
%undergoing larger-than-median tangential movements
%\cite{Kruyt:2007a}.
%
\item
The probability density of the normal contact forces usually
decreases exponentially for forces that are greater than
the mean 
%\citep{Mueth:1998a,Majmudar:2005a}.
\citep{Majmudar:2005a}.
With forces less than the mean, however, the density is more uniform
than exponential.
\item
Strength at the critical state increases with an increasing 
inter-granular friction
coefficient, but the relation is non-linear, and little
strength gain occurs when the friction coefficient increases beyond
0.3 
%\citep{Thornton:2000a,Suiker:2004a,daCruz:2005a,Kruyt:2006a,Hartl:2008a}.
\citep{Thornton:2000a,Hartl:2008a}.
Fabric anisotropy also increases with an increasing friction
%coefficient \citep{Cambou:2004a,daCruz:2005a,Kruyt:2014a}.
coefficient \citep{daCruz:2005a,Kruyt:2014a}.
\end{enumerate}
\begin{enumerate}[{C.}1.]
\item
Relatively few contacts attain the frictional
limit at any particular moment during flow,
and sliding typically occurs among only 8\%--20\% of the contacts
in assemblies of disks and spheres
\citep{Radjai:1998a,Thornton:2000a}.
The fraction of sliding contacts 
is reduced when the friction coefficient
is increased \citep{Thornton:2000a}.
\item
For contacts that are not sliding, the probability
density of the mobilized
friction is greatest for the nearly neutral condition of zero
tangential force, and the density decreases with increasing
mobilized friction \citep{Majmudar:2005a}.
%\item
%Tordesillas 3-cycles.
%
\end{enumerate}
\begin{enumerate}[{D.}1.]
\item
Internal force, movement, and deformation are highly heterogeneous
and are spatially organized across scales of ten or more particles
\citep{Kuhn:2003d}.
Contact forces are patterned in ``force chains''
of highly loaded particles that
are roughly aligned with the major principal compressive stress
\citep{Majmudar:2005a}.
Deformation is localized into obliquely oriented microbands
of thickness 1--3 particle diameters 
\citep[e.g.,][]{Kuhn:1999a}
and into shear bands
with a thickness of 8--20 diameters \citep{Desrues:2004a}.
Local stiffness \citep{Tordesillas:2011a} and local dilation and contraction 
are often clustered \citep{Kuhn:1999a}, and
chains of rapidly rotating particles are organized obliquely
to the major principal stress axes
\citep{Kuhn:1999a}.
The non-affine particle movements in 2D assemblies
are coordinated in large
circulating vorticity cells that encompass several dozens of
particles \citep{Williams:1997c}.
In 2D materials,
the topology of the contact network is also
spatially organized: voids surrounded
by 3-5 particles form elongated chains, and larger voids are
often organized as clusters \citep{Kuhn:2014a}.
\end{enumerate}
These and other characteristics, as determined from
new DEM results, will be discussed and compared in relation
to a proposed model of critical state flow.
Whereas the results listed above were 
gained from descriptive, observational studies, 
the current work is largely predictive.
The above characteristics are arranged as follows:
``A'' characteristics serve as the basis in developing the model,
``B'' characteristics are favorably predicted by the model,
``C'' characteristics are not well predicted or
must be forced into agreement with an \emph{ad hoc}
intervention,
% constraint on the model,
and ``D'' characteristics are inaccessible,
due to the nature of the model.
\par
The proposed model focuses on the movements, forces, and
orientations of inter-particle contacts during shear-driven flow.
The model is premised on the \emph{conjecture} that the
critical state is a condition of maximum micro-scale disorder
among these contact characteristics.
This maximally disorder state
is extracted,
%and this maximally disordered condition 
and its resulting conditions are shown to exhibit many of the
behaviors listed above.
%We will henceforth use the terms disorder and entropy
%interchangeably,
With some reluctance, we use the term ``entropy'' interchangeably with
disorder,
with entropy connoting
a statistical measure of micro-scale unpredictability (i.e., as in
\citealp{Shannon:1948a}) rather than an
extensive thermodynamic quantity having an intensive 
temperature-like dual
(in contrast with the compactivity or angoricity
settings, e.g. % of Edwards, Blumenfeld, and coworkers, e.g.
\citealp{Blumenfeld:2009a}).
%; also see \citealp{Scharle:1989a}
%for a discussion).
%
\par
Principles of maximum disorder have been applied
to granular materials for several decades.
%In these studies, disorder is usually expressed with the
%Shannon entropy\ \cite{Shannon:1948a,BenNaim:2008a}.
\citet{Brown:2000b} conducted
experiments on two-dimensional assemblies of spheres, 
and by applying
a back-and-forth shearing, found that
disorder in the local density and coordination
number increased with each shearing cycle. 
The Jaynes maximum entropy (MaxEnt) formalism
has typically been used to predict the condition of maximum disorder
\citep{Jaynes:1957a}.
%Maximum disorder (entropy) principles 
This approach has been applied 
to the local fabric of disk assemblies by categorizing 
voids into several canonical 
types \citep{Brown:2000a}.
Similar maximum entropy approaches have also been applied to
local packing 
%density \citep{Moroto:1983a,Edwards:1989a,Kumar:2005a,Yoon:2012a},
density \citep{Yoon:2012a},
to the contact forces
%\citep{Coppersmith:1996a,Edwards:2001a,Chakraborty:2010a},
\citep{Chakraborty:2010a},
to the contact orientations \citep{Troadec:2002a},
and to the contact displacements
and bulk elastic moduli \citep{Rothenburg:2009a}.
Among these aspects,
%of granular arrangement and behavior,
the distribution of contact force has recently received the
greatest attention.
%, with the focus primarily upon static,
%non-flowing granular assemblies.
Early theories derived probability densities
of force distribution by addressing the
bulk mean stress but without respecting the local force equilibrium
%\citep{Rothenburg:1980a,Blumenfeld:2003a,Bagi:2003b,Goddard:2004a}.
\citep{Goddard:2004a}.
More recent approaches
to the distribution of contact force
have enforced the local equilibrium of particles
and examined the statically admissible space of contact forces within a 
granular system 
%\citep{Coppersmith:1996a,Snoeijer:2004a,Metzger:2008a,Chakraborty:2010a}.
\citep{Chakraborty:2010a}.
These disorder theories address frictionless, non-flowing
assemblies and do not address the anisotropies of force, fabric,
and movement that accompany flow.
The paper not only admits flow 
but relies upon the shearing deformation
to drive the contact movements and the contact forces that generate the
bulk anisotropies of fabric and force.
\par
Our approach also differs from mean field theory, 
which is currently the dominant paradigm for relating the
micro- and macro-scale behaviors of dense materials, 
primarily at small strains.
Mean-field theories include
%the upscaling methods of \citep{Jenkins:1993a,Chang:1996a}, 
the upscaling methods of \citep{Jenkins:1993a}, 
in which the small-strain bulk stiffness is estimating
by assuming that the particle motions or contact forces around
a central particle conform to an affine, 
homogeneous kinematic (motion)
or static (stress) field.
We will not impose such affine restrictions, as doing so
would imply a strong order in the critical state condition --- an order
that simply does not exist \citep{Kuhn:2003d}.
Most mean-field approaches
%the surroundings of a particle might be described
%in a probabilistic manner, but the methods 
are also inherently deterministic insofar as they 
result in the unique
response of a particle for a given description of its neighborhood.
%when given particular locations of its neighbors.
\citet{Rothenburg:2009a} 
extended such methods by adopting
probability densities for both the surroundings and the response. %,
% and they established bounds on the small-strain stiffness moduli 
% by applying a maximum entropy principle.
The current work also adopts %probability densities both
%for the contact quantities and for the bulk response, 
a probabilistic framework,
but the approach is intended for granular flow in which the micro-scale
interactions are dominated by tangential contact motions.
%\par
%Just as kinetic theories of gas and of dilute, collisional granular flows
%address the distribution of molecular velocities and the relationships
%among state characteristics but without attending to
%details of individual molecules or collisions
%\citep[e.g.,][]{Dufty:2000a},
%the paper addresses only the
%bulk characteristics
%of stress, deformation, and dissipation that arise from
%distributions of certain contact quantities, 
%without respecting local equilibrium of the particles or the
%kinematic compatibility of their movements.
%In a sense, we will ignore the particles altogether
%and focus upon their contacts, which are assumed to be
%persistent (non-collisional) across small increments of bulk deformation.
\par
The plan of the paper is as follows.
We build a micro-scale flow model by first
identifying an essential set of contact quantities,
forming a rather comprehensive phase space of motion,
force, and arrangement
(Sec.~\ref{sec:variables}).
We regard these quantities as random variables with a
bulk probability density distribution (Sec.~\ref{sec:probabilities})
and apply constraints, both rational and empirical, to this distribution
(Secs.~\ref{sec:mean}--\ref{sec:eta}).
By themselves, these constraints do not describe a unique
distribution. %: multiple distributions are consistent
% with the constraints.
We assume that all such micro-states satisfying the
constraints are equiprobable during critical state flow and that
the most likely macro-state (i.e., density distribution) is the one that
encompasses the greatest breadth of micro-states, maximizing
the system's disorder.
A maximum entropy condition is
applied to the contact attributes to arrive at a predicted
distribution %corresponding to the largest Shannon entropy
(Sec.~\ref{sec:Entropy}).
Section~\ref{sec:Compare} describes the model's solution and compares it with
past observations and with
the results of new DEM simulations. %,
% with particular attention given to the model's predictions of
% anisotropies in the contact forces and movements.
\par
Although the principles presented in these sections are fundamental,
the set of constraints sparse,
and the results promising, 
their evaluation is computationally taxing.
Complete computational details, therefore, are provided in appendices.
We end the paper by considering
a number of questions raised by some 
difficulties and shortcomings of the entropy model:
which aspects of granular flow are amenable
to simple, sweeping statistical approaches, and which aspects
require closer attention to the details of grain interactions?
What additional constraining information is most appropriate
for these entropy methods?
How can a model that is conceptually simple be both
predictive and opaque?
%How can a 2D model of disks be extended to three dimensions and to
%different particle shapes?
%
%
\section{Entropy Model}
%We develop a model, based upon a maximum disorder
%principle, for estimating the probability distributions of
%contact forces (both normal and tangential), contact movements, 
%and contact orientations during slow, quasi-static 
%frictional critical
%state flow.
The model is developed in a 
two-dimensional setting of assemblies of disks that are
poly-disperse but of a narrow size range.
Because of the assumed small poly-dispersity, 
we can overlook the tendency of mono-disperse assemblies 
to develop ordered, crystallized (low entropy)
particle arrangements.
A narrow size range also permits characterizing the disk sizes by
their mean diameter $\overline{\ell}$.
The model is intended for durable and
nearly rigid disks, in which contact
indentations are small relative to particle size:
the contacts are assumed rigid-frictional and without contact moments.
%The first model takes a naive approach by applying a
%complete set of reasonable and general constraints, hence termed
%a ``complete'' model, with the only substantive input being
%the friction coefficient.
%This model is shown to be qualitatively consistent with
%many trends of critical state flow.
%A second model improves upon certain quantitative predictions,
%but only when afforded the advantage of additional and specific
%information
%that normally would not be readily available from physical experiments.
%
%\par
%Focusing on the contacts between particles rather than
%on the particles themselves,
%we begin by identifying the discrete contact quantities that
%characterize micro-states of a disk assembly.
%The prevalence of these quantities in a micro-state is expressed
%as a probability density distribution.
%We then establish constraints on the probabilities, as
%informed by certain stress, deformation, and dissipation conditions
%and with other information of an empirical origin.
%By themselves, these constraints do not describe a unique
%distribution: multiple distributions are consistent
%with the constraints.
%We assume that all micro-sates satisfying these
%constraints are equiprobable during critical state flow and that
%the most likely macro-state (i.e., density distribution) is the one that
%encompasses the greatest breadth of micro-states, maximizing
%the system's disorder.
%This assumption leads to a predicted probability density of the
%most likely macro-state during critical state flow.  
%Predictions are then compared with the results
%of DEM simulations.
%
\subsection{Contact quantities}\label{sec:variables}
The foundation of any entropy model is the set of
chosen quantities that characterize the micro-states that comprise the full
phase space of a physical system.
The quantities that one chooses will largely determine
the nature of the model and the phenomena that are addressed.
For example, a phase space composed exclusively of particle arrangements
will lead to a random isotropic state.
A phase space composed exclusively of contact forces will lead to estimates
of the static force distribution
\citep[e.g.,][]{Goddard:2004a}.
Our intention is to model conditions of movement, force, and arrangement
during granular flow, and we choose a rather large, comprehensive
phase space of certain \emph{contact quantities}. 
These quantitites
can be used
to extract or to constrain the bulk deformation rate tensor,
stress tensor, and fabric tensor,
and these micro-quantities are all accessible from simulations
of discrete particle systems.
% Particles vs. contacts XXXX.
%
\par
\begin{figure}
\centering
\includegraphics{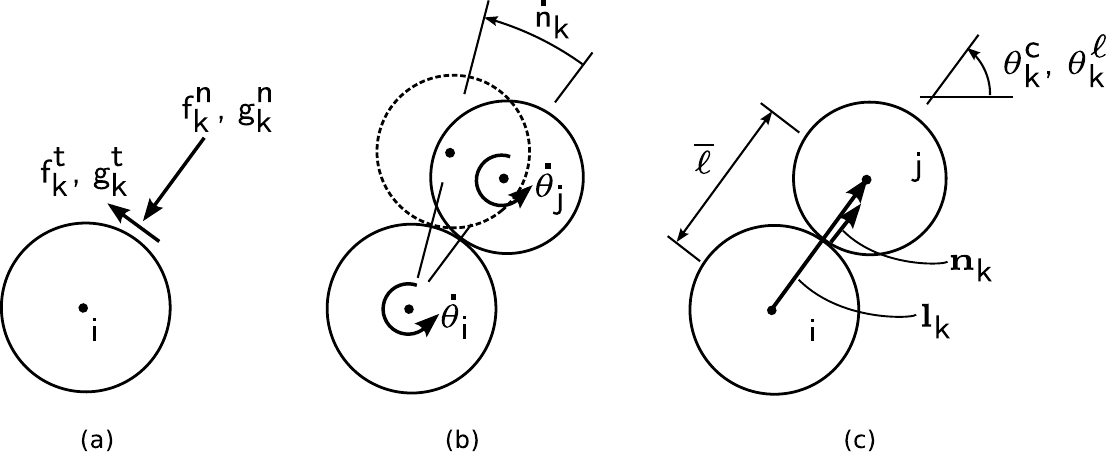}
\caption{Contact quantities of contact $k$ 
between two contacting disks $(i,j)_{k}$.
Direction conventions
for biaxial symmetry
are shown in Fig.~\ref{fig:Conventions}. \label{fig:TwoParticles}}
\end{figure}

A micro-state of a large two-dimensional assembly of disks
is enumerated
%by the values of
with six quantities at each of the
assembly's $M$ loading-bearing contacts, such that
a micro-state of the entire assembly is a single point within a
phase space of dimension $M^{6}$.
The six quantities, listed in Table~\ref{table:Quantities},
apply to a contact $k$ that is shared between
two particles, $i$ and $j$, which comprise an
ordered pair $(i,j)_{k}$ (Fig.~\ref{fig:TwoParticles}a).
\begin{table}
\caption{\small Contact quantities of the entropy model
         (Fig.~\ref{fig:TwoParticles})
         \label{table:Quantities}}
\centering\small
\begin{tabular}{ll}
\toprule
Quantity & Description\\
\hline
$g^{\text{n}} \in \mathbb{R}^{+}$ & 
  Compressive normal contact force,
  $f^{\text{n}}/(p_{\text{o}}\overline{\ell})$\\
%$g^{\text{t}} \in  [-\mu g^{\text{n}},\mu g^{\text{n}}]$ & 
%  $^{a}$Tangential contact force,
%  $f^{\text{t}}/(p_{\text{o}}\overline{\ell})$\\
$g^{\text{t}}$ &
  $^{a,b}$Tangential contact force,
  $f^{\text{t}}/(p_{\text{o}}\overline{\ell})$\\
$\theta^{\text{c}}\in [0,2\pi)$ &
  $^{a}$Orientation of contact normal vector $\mathbf{n}^{\text{c}}$\\
$\theta^{\ell}\in [0,2\pi)$ &
  $^{a}$Orientation of branch vector $\mathbf{l}^{\ell}$\\
$\dot{\phi}_{\text{slip}}\in\mathbb{R}$ & $^{a,b}$%
  $\sqrt{2/3}$ times contact slip, ``$\text{slip}_{k}$''\\
$\dot{\phi}_{\text{rigid}}\in\mathbb{R}$ &
  $^{a}$$\sqrt{3}$ times rigid contact movement, ``$\text{rigid}_{k}$''\\
\midrule
\multicolumn{2}{l}{\raggedright%
  $^{a}$Directional conventions are shown in Fig.~\ref{fig:Conventions}.}\\
\multicolumn{2}{l}{%
  $^{b}$$g^{\text{t}}$ and $\dot{\phi}_{\text{slip}}$ are 
    also limited by Eq.~(\ref{eq:GtRestrict}).}\\
\bottomrule
\end{tabular}
\end{table}
Contacts are assumed to be 
enduring, non-collisional, and persistent over
the brief interval for which bulk averages are extracted.
Note that, henceforth, we will largely ignore the association of
contacts with particles, thus overlooking
the underlying topology of the contact network and relinquishing an ability
to assure local equilibrium or kinematic compatibility.
The compressive normal contact force 
$f^{\text{n}}_{k}$ acts at the contact ``$k$'' 
upon particle ``$i$'' (Fig.~\ref{fig:TwoParticles}b).
Because the average normal force in a two-dimensional assembly
is roughly proportional to the 
mean diameter $\overline{\ell}$ and to the mean stress $p_{\text{o}}$,
%(having units $\text{force}\cdot\text{length}^{-1}$),
we normalize the individual normal forces as 
$g^{\text{n}}_{k}=f^{\text{n}}_{k}/(p_{\text{o}}\overline{\ell})$.
The scalar tangential force on $i$ is $f^{\text{t}}_{k}$, normalized
as $g^{\text{t}}_{k}=f^{\text{t}}_{k}/(p_{\text{o}}\overline{\ell})$,
%directed counterclockwise around $i$.
and this force component is limited by friction,
as characterized by coefficient $\mu$:
$|g^{\text{t}}_{k}| \leq \mu g^{\text{n}}_{k}$.
% Henceforth, the tangential force will be further reduced
% to the range $[-1,1]$ and expressed as the contact quantity $\psi_{k}$,
% with $\psi_{k}=f^{\text{t}}_{k}/(\mu f^{\text{n}}_{k})
% =g^{\text{t}}_{k}/(\mu g^{\text{n}}_{k})$, with the values $-1$ and $1$
% representing the friction limit
% and intermediate values representing the fraction of
% mobilized friction.
The sign convention associated with $f^{\text{t}}$ and
$g^{\text{t}}$ will be described
further below.
\par
Only tangential contact movements are modeled,
%exclusively within the model,
since normal motions (those that alter contact indentations
or cause the separation or creation of contacts) are known to
contribute negligibly to bulk deformation during flow
(characteristics A.1 and A.2 in the Introduction).
During deformation, particles will both rotate and shift relative to each other, 
and we adopt two systems for describing these movements.
In the first system,
scalar $\dot{n}_{k}$ represents an angular rate produced by 
\emph{translation}
of the center of
$j$ relative to the center of $i$ --- a
rotational rate of the unit normal vector
$\mathbf{n}_{k}$ directed outward 
from $i$ (Fig.~\ref{fig:TwoParticles}c).
Note that the rate vector $\dot{\mathbf{n}}_{k}$ is orthogonal to
$\mathbf{n}_{k}$.
The relative motions of the particles at their contact are produced
both by this tangential shifting of the particles'
centers (i.e., the velocity $\overline{\ell}\dot{n}_{k}$)
and by the two particles' rotations, 
$\dot{\theta}_{i}$ and $\dot{\theta}_{j}$
(Fig.~\ref{fig:TwoParticles}c).
Motions in the first system are described by the
quantities $\dot{n}_{k}$, $\dot{\theta}_{i}$, and $\dot{\theta}_{j}$.
The second system more directly addresses the contact interactions,
as expressed through three mechanisms:
1) slip (sliding) between the particles,
2) rolling movements that produce no slip,
and 3) rigid rotations of the particle pair, also producing no slip.
For equal-size disks of diameter $\overline{\ell}$, the mechanisms
are as follows \citep{Kuhn:2005c}:
\begin{align}
\text{slip}_{k} &= 
\dot{n}_{k} 
  - \textstyle{\frac{1}{2}}(\dot{\theta}_{i}+\dot{\theta}_{j})\notag\\
\text{roll}_{k} &=
  \dot{\theta}_{i}-\dot{\theta}_{j}\label{eq:slipk}\\
\text{rigid}_{k} &=
  \textstyle{\frac{1}{3}}(\dot{n}+\dot{\theta}_{i}+\dot{\theta}_{j})
  \notag
\end{align}
all having the units time$^{-1}$.
Because all three mechanism usually occur simultaneously at any contact,
we decompose the motions
$\dot{n}_{k}$, $\dot{\theta}_{i}$, and $\dot{\theta}_{j}$ as
\begin{equation}
\begin{bmatrix}
\dot{\phi}_{\text{slip},k}\\
\dot{\phi}_{\text{roll},k}\\
\dot{\phi}_{\text{rigid},k}
\end{bmatrix}
=
\begin{bmatrix}
\sqrt{2/3} & -\sqrt{1/6} & -\sqrt{1/6}\\
0 & \sqrt{1/2} & -\sqrt{1/2}\\
\sqrt{1/3} & \sqrt{1/3} & \sqrt{1/3}
\end{bmatrix}
\begin{bmatrix}
\dot{n}_{k}\\
\dot{\theta}_{i}\\
\dot{\theta}_{j}
\end{bmatrix}\label{eq:PhiRates}
\end{equation}
forming an orthogonal separation of particle movements into
three ``$\dot{\phi}$'' contact rates.
%also having the units $\text{time}^{-1}$.
Two quantities,
%Constraints will later be placed upon certain averages of
$\dot{n}_{k}$ and ``$\text{slip}_{k}$'',
are most relevant in the model,
as they determine the bulk deformation and dissipation rates.
Because both quantities can
be expressed as linear combinations of $\dot{\phi}_{\text{slip},k}$
and $\dot{\phi}_{\text{rigid},k}$, the model does not require
$\dot{\phi}_{\text{roll},k}$,
%As such, only movements $\dot{\phi}_{\text{slip},k}$ 
%and $\dot{\phi}_{\text{rigid},k}$ will be included,
%treated as fundamental contact quantities in the model,
and the reduction of three quantities to two
greatly reduces computational demands.
%
%Assuming the two particle are nearly the same size,
%we introduce the quantity 
%$\dot{\phi}_{k}=\frac{1}{2}(\dot{\theta}_{i}+\dot{\theta}_{j})$ to
%capture the effect of the two particles'
%rotations on the relative contact motion, 
%such that the full
%frictional slip between two rigid particles (i.e., the movement 
%of a material point attached to $j$
%relative to its counterpart on $i$) is given by
%
%\begin{equation}
%\text{slip}_{k} = 
%\overline{\ell} \left( \dot{n}_{k} - \dot{\phi}_{k} \right) \label{eq:slip}
%\end{equation}
%
%Note that the two quantities $\dot{n}$ and $\dot{\phi}$ do not
%fully describe the two particles' motions,
%%as the particles can also roll at their contact without sliding
%(i.e. with a motion $\dot{\theta}_{i}-\dot{\theta}_{j}$), but we will
%not consider this rolling aspect in the paper.
%
\par
As the final contact quantities,
we introduce two orientation angles,
$\theta^{\text{c}}$ and $\theta^{\ell}$,
with angle $\theta^{\text{c}}$ giving the direction
of the contact normal $\mathbf{n}_{k}$, and angle
$\theta^{\ell}$ representing the direction
of the branch vector $\mathbf{l}_{k}$, 
of length $\overline{\ell}$, that joins
the center of $i$ to that of $j$ (Fig.~\ref{fig:TwoParticles}a).
Although the two angles are identical for disks, they
are distinguished because of their different roles
in expressing stress and strain within the assembly.
Both angles are measured from particle $i$.
\par
\begin{figure}
\centering
\includegraphics{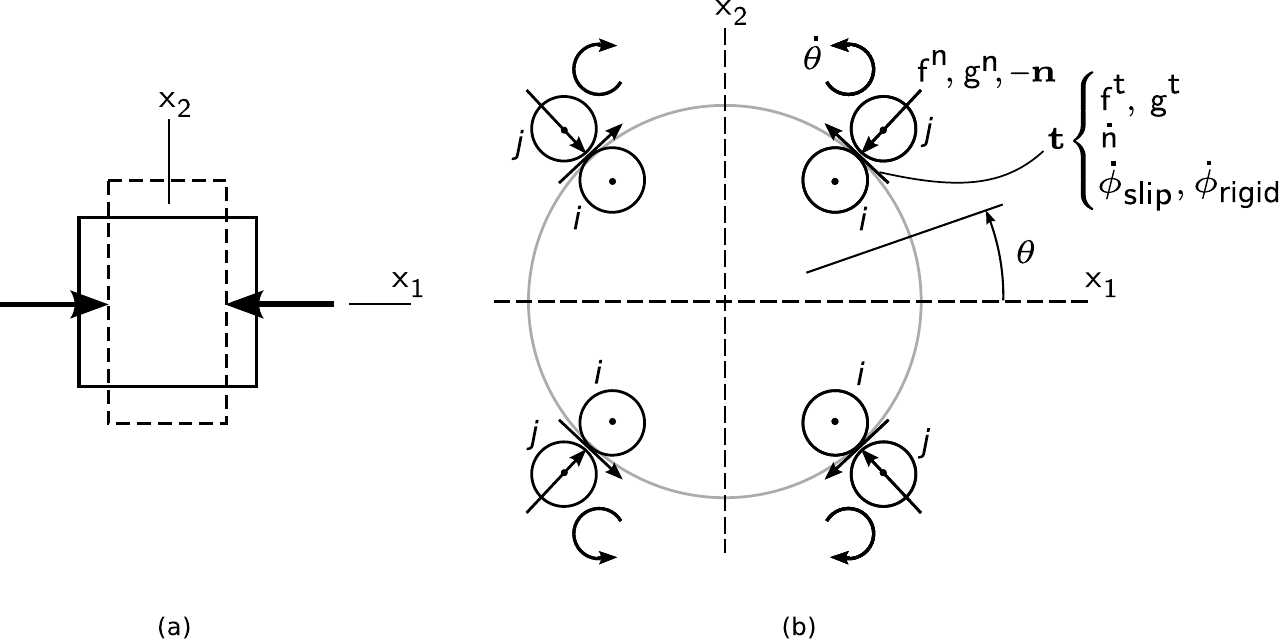}
\caption{Conventions for unit vectors $\mathbf{n}$ and $\mathbf{t}$
and rotations $\dot{\theta}$
when analyzing horizontal biaxial compression. 
``Forward'' directions are shown. \label{fig:Conventions}}
\end{figure}

Although the model is quite general, it will be applied to the particular
condition of biaxial compression, in which the $x_{1}$ width of
the assembly is reduced while the $x_{2}$ height expands
(Fig.~\ref{fig:Conventions}a).
We adopt a sign convention that takes advantage of the symmetry of
these conditions and assists bookkeeping
when the model is later compared with DEM results.
The convention, shown in Fig.~\ref{fig:Conventions}b, 
assigns a direction to the
tangential unit vector $\mathbf{t}_{k}$ corresponding
to the sliding that would be expected
if the particle motions roughly
conformed to affine deformation
during horizontal biaxial compression.
This direction --- clockwise or counterclockwise --- alternates
across quadrants of $\theta$.
Unit vector $\mathbf{t}_{k}$ applies to the (scalar) 
tangential force on $i$ (i.e., $f^{\text{t}}_{k}$ and
$g^{\text{t}}_{k}$) 
and to the relative contact movements
($\dot{\theta}$, $\dot{\phi}_{\text{slip},k}$, etc.).
Positive values of $f^{\text{t}}_{k}$,
$\dot{\phi}_{\text{slip},k}$, etc.,
which are aligned with $\mathbf{t}_{k}$,
are called ``forward'' forces and movements; 
whereas, negative values are in the ``reverse''
direction $-\mathbf{t}_{k}$.
Normal force $f^{\text{n}}_{k}$ is always compressive, and
unit vector $\mathbf{n}_{k}$ is directed outward from particle~$i$.
\par
The domains of the six quantities are shown in
Table~\ref{table:Quantities}.
We also place unilateral and rigid-frictional \emph{constitutive} limits upon the
model,
\begin{align}
g^{\text{n}}_{k} &\in \mathbb{R}^{+} \label{eq:GnRestrict}\\
g^{\text{t}}_{k} &\in %(-\mu g^{\text{n}}_{k},\mu g^{\text{n}}_{k})
   %\text{ and } g^{\text{t}}_{k} \in
   \begin{cases} 
    -\mu g^{\text{n}}_{k} & 
     \Leftrightarrow\quad \dot{\phi}_{\text{slip},k}<0 \\
     (-\mu g^{\text{n}}_{k},\mu g^{\text{n}}_{k}) & 
     \Leftrightarrow\quad
     \dot{\phi}_{\text{slip},k}=0\\
     \mu g^{\text{n}}_{k} &
     \Leftrightarrow\quad
     \dot{\phi}_{\text{slip},k}>0
   \end{cases} \label{eq:GtRestrict}
\end{align}
further restricting the domains
of $g^{\text{n}}_{k}$, $g_{k}^{\text{t}}$,
and $\dot{\phi}_{\text{slip},k}$.
These restrictions mean
that the particles are unbonded ($f^{\text{n}}_{k},g^{\text{n}}_{k}>0$), 
and the tangential force $f^{\text{t}}_{k}$ is limited by friction to
the range $[-\mu f^{\text{n}}_{k},\,\mu f^{\text{n}}_{k}]$.
Positive (forward) slip, $\dot{\phi}_{\text{slip},k}>0$, is only
admissible with a corresponding positive tangential force,
$f^{\text{t}}_{k}=\mu f^{\text{n}}_{k}$; 
whereas, negative (reverse) slip
is only possible for the opposite condition,
$f^{\text{t}}_{k}=-\mu f^{\text{n}}_{k}$.
When the friction limit is not reached, 
$|f^{\text{t}}_{k}|<\mu f^{\text{t}}_{k}$, the slip is zero:
$\dot{\phi}_{\text{slip},k}=0$.
In adopting Eqs.~(\ref{eq:GnRestrict}) and~(\ref{eq:GtRestrict}),
we forego a more elaborate elastic-frictional model,
since critical state flow is insensitive
to contact elasticity 
(characteristic A.\ref{item:ContactElasticity} of the Introduction).
Note that the model is rate-independent, since forces
$f^{\text{t}}_{k}$ depend only on the directions of
$\dot{\phi}_{\text{slip},k}$ and not on their magnitudes.
\subsection{Probabilities and constraints}\label{sec:probabilities}
In the model, 
we do not individualize the six ``$k$'' quantities
among the $M$ contacts (a phase space of dimension $M^{6}$) but instead
treat the quantities as continuous random variables with probability density
$p(\cdots)$:
\begin{equation}
p(\cdots) = p( g^{\text{n}},g^{\text{t}}, 
        \theta^{\text{c}},\theta^{\ell},
        \dot{\phi}_{\text{slip}},\dot{\phi}_{\text{rigid}})
\end{equation}
dropping the $k$ subscripts.
Although we write the complete phase space of possible
\emph{micro-states} as 
$\{ g^{\text{n}}_{k},g^{\text{t}},\theta^{\text{c}}_{k},
\theta^{\ell}_{k},
\dot{\phi}_{\text{slip},k},\dot{\phi}_{\text{rigid},k}\}$,
%or simply $\{\cdots\}$.
all micro-states within this $M^{6}$ space that
share the same probability density $p(\cdots )$ 
comprise a common \emph{macro-state},
an element in a subspace of $\mathbb{R}^{6}$ and
written without parentheses:
$( g^{\text{n}},g^{\text{t}}, 
        \theta^{\text{c}},\theta^{\ell},
        \dot{\phi}_{\text{slip}},\dot{\phi}_{\text{rigid}})$
        or $(\cdots)$.
As will be seen, density $p$ contains comprehensive
information about bulk quantities, such as stress and fabric,
and about the correlations among contact 
orientation, force, and movement.
However, by turning from individual ``$k$'' 
contact quantities to their
gross probability distribution, we forfeit the possibility
of tracking individual contacts.
%We also lose the ability to discern any spatial localization
%or meso-scale patterning of the movements and forces.
In return, we gain a certain economy of expression for 
predicting overall distributions of
the six quantities and the statistical relationships among them.
\par
%In the model of critical state flow,
A wide range of the six quantities are observed
in experiments and simulations, and correlations among the
quantities are found during critical state flow, as
enumerated in the Introduction.
These correlations are expressions of order, which we consider
a consequence of certain constraints that link the micro and
bulk behaviors.
We will restrict the admissible
macro-states with five constraints: 
four fundamental constraints and one
auxiliary constraint.
These constraints bring \emph{a priori} information
that will bias the distribution $p(\cdots )$.
Each ``$i$''th constraint is expressed with a constraining function 
$\Gamma_{i}(\cdots)$ of the six random variables,
\begin{equation}
\Gamma_{i} 
  ( g^{\text{n}},g^{\text{t}}, 
        \theta^{\text{c}},\theta^{\ell},
        \dot{\phi}_{\text{slip}},\dot{\phi}_{\text{rigid}})
\quad\text{or simply}\quad
\Gamma_{i}(\cdots )
\end{equation}
The expected value $\langle \Gamma_{i}\rangle$ of the
function
is found through 6-fold integration across the full phase space,
with each random variable integrated over the range given in
Table~\ref{table:Quantities}:
\begin{equation}
\left\langle \Gamma_{i}(\cdots) \right\rangle = 
\underset{( g^{\text{n}},g^{\text{t}}, 
        \theta^{\text{c}},\theta^{\ell},
        \dot{\phi}_{\text{slip}},\dot{\phi}_{\text{rigid}} )}
        {\idotsint} 
\Gamma_{i}(\cdots) \,p(\cdots)\label{eq:GammaInt}
\end{equation}
and measure $dg^{\text{n}}dg^{\text{t}}
        d\theta^{\text{c}}d\theta^{\ell}
        d\dot{\phi}_{\text{slip}}d\dot{\phi}_{\text{rigid}}$
is implied in these integrations.
The integrations are described in greater detail
in \ref{sec:Integrations}.
%(these integrations are described in Section~\ref{sec:Integrations}),
%writing the dimension-6 domain as
%$( g^{\text{n}},\psi,\theta^{\text{c}},\theta^{\ell},
%\dot{\phi}_{\text{slip}},\dot{\phi}_{\text{rigid}} )$
%or simply $(\cdots )$
%without the ``$k$'' subscripts.

\par
%Constraints are effected as follows.
The expected values of functions $\Gamma_{i}(\cdots )$ will be constrained to
certain average values,
\begin{equation}
\text{Constraint } i \; \Rightarrow \;
\langle \Gamma_{i}(\cdots)\rangle = \overline{\Gamma}_{i}
\label{eq:Constraints}
\end{equation}
with the prescribed averages, $\overline{\Gamma}_{i}$, 
developed below.
That is, each restriction (\ref{eq:Constraints})
is a moment constraint on density $p(\cdots)$.
\par
An essential ``zeroth'' constraint follows from the certainty
that the six quantities lie within their full ranges: %, as
%expressed in three equations,
%
\begin{equation}
\begin{gathered}
\left\langle\Gamma_{0}(\cdots)\right\rangle = \overline{\Gamma}_{0}
\;\text{, where}
\\
\Gamma_{0}(\cdots)=1
\quad\text{and}\quad
\overline{\Gamma}_{0} = 1
\end{gathered}\label{eq:Certainty}
\end{equation}
such that integration of $p(\cdots )$ 
over its full domain equals 1. 
%
%\subsection{The complete model}\label{sec:complete}
%
\par
%The model of dense flow under biaxial conditions
%is defined by seven constraints.
The four fundamental constraints
express general principles of isochoric dissipative flow
at constant stress and are derived in 
Sections~\ref{sec:mean}--\ref{sec:auxiliary}.
%One auxiliary constraint applies 
%to the particular case of biaxial loading
%(Section~\ref{sec:auxiliary}), and
The auxiliary constraint applies additional information
gained from DEM simulations (Section~\ref{sec:eta}).
The constraints are summarized in 
Table~\ref{table:Constraints1}.
\begin{table}
\centering\small
\caption{\small Summary of constraints in entropy model.
          \label{table:Constraints1}}
\begin{tabular}{r@{\ }lllc}
\toprule
Constraint & $i$ &
  Description & Variables in $\Gamma_{i}(\cdots)$ & Eq.\\
\midrule
%Essential & 0 & $\int p(\cdots)=1$ & 
%  None & (\ref{eq:Certainty})\\
\rule{0em}{1.2em}Fundamental & 1 &
  Mean stress $=p_{\text{o}}$ &
  $(g^{\text{n}},\cdot,\cdot,\cdot,\cdot,\cdot)$ &
  (\ref{eq:Gamma2})\\
& 2 &
  Dissipation consistency &
  $(g^{\text{n}},g^{\text{t}} ,\theta^{\text{c}},\cdot,
    \dot{\phi}_{\text{slip}},\cdot)$ &
  (\ref{eq:Gamma3})\\
& 3 & 
  Isochoric flow& 
  $(\cdot,\cdot,\theta^{\text{c}},\theta^{\ell},\dot{\phi}_{\text{slip}},
    \dot{\phi}_{\text{rigid}})$ & 
  (\ref{eq:Gamma1})\\
& 4 & Biaxial rate, $\dot{\varepsilon}$ &
  $(\cdot,\cdot,\theta^{\text{c}},\theta^{\ell},\dot{\phi}_{\text{slip}},
    \dot{\phi}_{\text{rigid}})$ & 
  (\ref{eq:Gamma4})\\
%          & 5 & Dispersion of slip movements, $\dot{\phi}_{\text{slip}}$ &
%  $(\cdot,\cdot,\cdot,\cdot,\dot{\phi}_{\text{slip}},\cdot)$ & 
%  (\ref{eq:Gamma5})\\
%          & 6 & Relative slip and rigid movements &
%  $(\cdot,\cdot,\cdot,\cdot,\dot{\phi}_{\text{slip}},
%    \dot{\phi}_{\text{rigid}})$ & 
%  (\ref{eq:Gamma6})\\
Auxiliary & 5 & Fraction of sliding contacts, $\eta$ &
  $(g^{\text{n}},g^{\text{t}},\cdot,\cdot,\cdot,\cdot)$ & 
  (\ref{eq:Gamma7})\\
\bottomrule
\end{tabular}
\end{table}
\subsection{Fundamental constraint~1: Constant mean stress}\label{sec:mean}
We model steady state flow under constant mean stress $p_{\text{o}}$,
a condition commonly applied in geotechnical testing.
The Cauchy stress in a large granular assembly is the volume average
of the contact dyads 
$\mathbf{l}^{\ell}_{k}\otimes\mathbf{f}^{\text{c}}_{k}$, 
\begin{equation}
\boldsymbol{\sigma} = 
\frac{1}{A} \sum_{k=1}^{M^{\text{int}}} 
\mathbf{l}_{k}^{\ell}\otimes\mathbf{f}_{k}^{\text{c}}
\label{eq:SumStress}
\end{equation}
in which
branch vector $\mathbf{l}_{k}^{\ell}$ joins particle $i$ to $j$,
contact force $\mathbf{f}_{k}^{\text{c}}$ acts upon $i$,
and $M^{\text{int}}$ is the number of contacts within the interior
of the two-dimensional region of area $A$
%\citep{Rothenburg:1980a,Christoffersen:1981a}.
\citep{Rothenburg:1980a}.
The branch vector length is approximated as $\overline{\ell}$, so that
\begin{equation}
\mathbf{l}_{k}^{\ell} \approx
\overline{\ell}\mathbf{n}_{k}^{\ell} \label{eq:lkc}
\end{equation}
and the contact force is the
sum of its tangential and (compressive) normal components:
\begin{align}
\mathbf{f}_{k}^{\text{c}} &=
-f^{\text{n}}_{k} \mathbf{n}_{k}^{\text{c}}
+ f^{\text{t}}_{k}\mathbf{t}_{k}^{\text{c}}\\
&=\overline{\ell}\,p_{\text{o}}
\left( -g_{k}^{\text{n}}\mathbf{n}_{k}^{\text{c}}
+ g_{k}^{\text{t}}\mathbf{t}_{k}^{\text{c}} \right)
% = \overline{\ell}\,p_{\text{o}}\,g_{k}^{\text{n}}
% \left( -\mathbf{n}_{k}^{\text{c}}+\psi\mu\mathbf{t}_{k}^{\text{c}}\right)
\label{eq:fkc}
\end{align}
In these and future expressions,
$\mathbf{n}^{\text{c}}$ %(or $\mathbf{n}^{p}$)
and $\mathbf{n}^{\ell}$ %(or $\mathbf{n}^{q}$) 
are unit orientation vectors
associated with the contact normal and branch vectors:
\begin{equation}
\mathbf{n}^{\text{c}} = 
  \begin{bmatrix}\cos\theta^{\text{c}}\\ \sin\theta^{\text{c}} \end{bmatrix}
  ,\quad
\mathbf{n}^{\ell} =
  \begin{bmatrix}\cos\theta^{\ell}\\ \sin\theta^{\ell}
  \end{bmatrix}\label{eq:N}
\end{equation}
For the particular sign convention %of $\mathbf{t}$ 
shown in
Fig.~\ref{fig:Conventions}b, unit vector $\mathbf{t}^{\text{c}}$
is
\begin{equation}
\mathbf{t}^{\text{c}} = 
K_{1}(\theta^{\text{c}})
\begin{bmatrix} -\sin\theta^{\text{c}}\\ \cos\theta^{\text{c}}\end{bmatrix},
\quad
K_{1}(\theta^{\text{c}})=
\operatorname{sgn}(\cos\theta^{\text{c}}\sin\theta^{\text{c}})
\label{eq:tc}
\end{equation}
where the signum function
$\text{sgn}()\in\{-1,1\}$
alternates in sign between coordinate quadrants.
\par
We replace the sum in Eq.~(\ref{eq:SumStress}) with an integration by 
substituting Eqs.~(\ref{eq:lkc}) and~(\ref{eq:fkc}),
multiplying by
probability $p(\cdots)$ and
the number of contacts $M^{\text{int}}$,
and integrating
across the full domain of the six quantities $(\cdots )$:
\begin{align}
\boldsymbol{\sigma} &= 
\underset{( g^{\text{n}},g^{\text{t}}, 
        \theta^{\text{c}},\theta^{\ell},
        \dot{\phi}_{\text{slip}},\dot{\phi}_{\text{rigid}} )}
        {\idotsint} 
\boldsymbol{\Gamma}_{\boldsymbol{\sigma}}(\cdots) p(\cdots ) \label{eq:IntSigma}  \\
\boldsymbol{\Gamma}_{\boldsymbol{\sigma}}
&=
\left( M^{\text{int}}\overline{\ell}^{2}\,p_{\text{o}} / A\right)\,
K_{2}( \theta^{\text{c}},\theta^{\ell})
\,
\mathbf{n}^{\ell}\otimes
( -g^{\text{n}}\,\mathbf{n}^{\text{c}} 
+ g^{\text{t}}\,\mathbf{t}^{\text{c}})
\label{eq:GammaSigma}
\end{align}
In this expression, we include a kernel function $K_{2}$, which is simply
the Dirac operator,
\begin{equation}
K_{2}( \theta^{\text{c}},\theta^{\ell}) = 
\delta(\theta^{\text{c}}-\theta^{\ell}) \label{eq:K1}
\end{equation}
The Dirac kernel formally dictates 
the correspondence of the unit branch vector and
the contact force vector for circular particles (by extension, unit vectors 
$\mathbf{n}^{\ell}$ and $\mathbf{n}^{\text{c}}$), 
as in the single sum of Eq.~(\ref{eq:SumStress}).
% This kernel is unlike $K_{1}$ in Eqs.~(\ref{eq:GammaL}) and~(\ref{eq:K2}),
% which effects the double integration of Eq.~(\ref{eq:IntegralL}) 
% and in which the two values
% $s^{p}$ and $s^{q}$ 
% (or angles $\theta^{\text{c}}$ and $\theta^{\ell}$) correspond to
% different points on the boundary of a granular assembly.
\par
Equations~(\ref{eq:IntSigma})--(\ref{eq:K1}) will later be used
to extract the predicted stress tensor during critical state flow,
but we now use them to enforce the first fundamental constraint
on the probability density $p(\cdots)$:
the negative
mean stress $-\frac{1}{2}\text{tr}(\boldsymbol{\sigma})$ must equal
the assigned pressure $p_{\text{o}}$.
The negative mean value of tensor function 
$\boldsymbol{\Gamma}_{\boldsymbol{\sigma}}$ is
\begin{equation}
-\frac{1}{2}\text{tr}\left( \boldsymbol{\Gamma}_{\boldsymbol{\sigma}} \right)
= \frac{1}{2}\frac{ M^{\text{int}}\overline{\ell}^{2}p_{\text{o}} }{A}\,
K_{2}( \theta^{\text{c}},\theta^{\ell})
\,g^{\text{n}} \label{eq:TraceSigma}
\end{equation}
and equating its integration in Eq.~(\ref{eq:IntSigma})
with $p_{\text{o}}$ leads to
the first fundamental constraint
\begin{equation}
\begin{gathered}
\left\langle\Gamma_{1}(\cdots)\right\rangle = \overline{\Gamma}_{1}
\;\text{, where}
\\
\Gamma_{1}(\cdots) = 
  K_{2}( \theta^{\text{c}},\theta^{\ell})\, g^{\text{n}}
\quad\text{and}\quad
\overline{\Gamma}_{1} = 2
\left( M^{\text{int}}\overline{\ell}^{2} / A \right)^{-1}
\end{gathered} \label{eq:Gamma2}
\end{equation}
in which we have divided by the leading constant 
in Eq.~(\ref{eq:TraceSigma}).
Equation~(\ref{eq:Gamma2})
simply requires the dimensionless mean normal force 
$\langle g^{\text{n}}\rangle$ 
(i.e., $\langle f^{\text{n}}\rangle/\overline{\ell}p_{\text{o}}$) 
to equal
the dimensionless quantity $\overline{\Gamma}_{1}$,
thus enforcing pressure $p_{\text{o}}$.
The contact density~$M^{\text{int}}\overline{\ell}^{2}/A$
is about 1.5, a value that can be found with DEM simulations
or estimated from characteristic A.4 of the Introduction.
% and be measured in experiments or DEM simulations
% and is about 1.5 (Sections~\ref{} and~\ref{}).
% in \ref{sec:Counting},
% by applying characteristic A.5 of the Introduction.
%
\subsection{Fundamental constraint~2: Dissipation consistency}\label{sec:dissipation}
During critical state flow at constant stress and volume, 
deformation is entirely plastic: all work done by
the applied stress is dissipated through internal
irreversible processes.
% Although some energy loss might result from the plastic
% deformation and fracture of particles,
% viscous dissipation 
% (sometimes modeled with a coefficient of restitution)
% is minimal for slow quasi-static flow, 
% and 
We assume that
the only available dissipation mechanism is frictional sliding
between particles.
This assumption is the basis of the second fundamental
constraint on the probability density $p(\cdots)$ and provides
the essential link among contact motions,
contact forces, bulk deformation, and bulk stress.
This constraint drives the anisotropies in force and movement.
\par
The frictional dissipation rate at the contact of two
disks is the product of its tangential contact force
$f^{\text{t}}_{k}=\overline{\ell}p_{\text{o}}g^{\text{t}}_{k}$
% =\overline{\ell}p_{\text{o}}(\psi_{k}\mu g^{\text{n}}_{k})$ 
and the relative slip of
the two particles' surfaces at their contact: 
the ``$\text{slip}_{k}$'' rate in Eq.~(\ref{eq:slipk}$_{1}$).
Applying Eq.~(\ref{eq:PhiRates}),
this slip rate corresponds to $\dot{\phi}_{\text{slip}}$ as
\begin{equation}
\text{slip}_{k} = \sqrt{3/2}\,\dot{\phi}_{\text{slip},k}
\end{equation}
%
%defined as
%$\overline{\ell}(\dot{n}_{k} - \dot{\phi}_{k})$ in 
%Eq.~(\ref{eq:slip}),
Noting
that $g^{\text{t}}$ and $\dot{\phi}_{\text{slip},k}$
must conform to the constitutive rigid-frictional restrictions of
Eqs.~(\ref{eq:GnRestrict}) and~(\ref{eq:GtRestrict}),
the frictional dissipation rate per unit of area is
\begin{equation}\label{eq:Dissip}
\text{Dissipation} = \frac{\overline{\ell}^{2}p_{\text{o}}}
     {A}
  \sum_{k=1}^{M^{\text{int}}}
  \sqrt{3/2}\,
  g^{\text{t}}
% \psi_{k}\,\mu\,g^{\text{n}}_{k}
  \,\dot{\phi}_{\text{slip},k}
\end{equation}
or, in terms of probability density $p(\cdots )$,
\begin{align}
\text{Dissipation} &=
\underset{( g^{\text{n}},g^{\text{t}}, 
        \theta^{\text{c}},\theta^{\ell},
        \dot{\phi}_{\text{slip}},\dot{\phi}_{\text{rigid}} )}
        {\idotsint}
\Gamma_{\text{f}}(\cdots) \,p(\cdots ) \label{eq:IntDissipation}\\
\Gamma_{\text{f}} &=
\frac{M^{\text{int}}\overline{\ell}^{2}\,p_{\text{o}}}{A}\,
\sqrt{3/2}\,
K_{2}( \theta^{\text{c}},\theta^{\ell})
\,g^{\text{t}}\,\dot{\phi}_{\text{slip}}
%g^{\text{n}} \left(\dot{n} - \dot{\phi}\right)
\label{eq:Gammaf}
\end{align}
where the Dirac kernel $K_{2}$ is applied again,
enforcing the coincidence of the contact forces
%$g^{\text{t}}$ 
(associated with contact orientations
$\theta^{\text{c}}$) and the contact motions
%$\overline{\ell}(\dot{n} - \dot{\phi})$
(associated with branch vector orientations $\theta^{\ell}$)
in the single sum of Eq.~(\ref{eq:Dissip}).
\par
The internal work rate of the Cauchy stress $\boldsymbol{\sigma}$
is the inner product
\begin{equation}
\text{Work}= \boldsymbol{\sigma}:\mathbf{D} \label{eq:Work}
\end{equation}
where $\mathbf{D}$ is the symmetric part of velocity gradient $\mathbf{L}$.
Although Eq.~(\ref{eq:IntSigma}) is an expression of
$\boldsymbol{\sigma}$, and an expression
for $\mathbf{L}$ is derived in the next section
(as Eq.~\ref{eq:IntegerL2}),
% expressions for both $\mathbf{L}$ and $\boldsymbol{\sigma}$
% are given in Eqs.~(\ref{eq:IntegerL2}) and~(\ref{eq:IntSigma}), 
directly substituting these expressions
into the full inner product of Eq.~(\ref{eq:Work}) will lead to
a non-linear constraint on probability $p(\cdots )$, making
its evaluation intractable.
The situation is greatly simplified when deformation can be 
expressed with a single non-zero parameter.
We consider the case of biaxial compression,
for which $\mathbf{D}$ is reduced to
\begin{equation}
\mathbf{D} = 
  \begin{bmatrix}
    -\dot{\varepsilon} & 0 \\ 0 & \dot{\varepsilon}
  \end{bmatrix}
\label{eq:Dmatrix}
\end{equation}
and the work rate is $-\dot{\varepsilon}(\sigma_{11}-\sigma_{22})$.
\par
Equating the rates of frictional dissipation and internal work in
Eqs.~(\ref{eq:IntDissipation}) and~(\ref{eq:Work})
and substituting Eqs.~(\ref{eq:IntSigma}),
(\ref{eq:GammaSigma}), and~(\ref{eq:Dmatrix})
lead to the second fundamental constraint on
density $p(\cdots )$ for critical state flow in
biaxial compression:
\begin{equation}
\begin{gathered}
\left\langle\Gamma_{2}(\cdots)\right\rangle = \overline{\Gamma}_{2}
\;\text{, where}
\\
\Gamma_{2}(\cdots ) = 
%\left( \frac{M^{\text{int}}\overline{\ell}^{2}p_{\text{o}}}{A}\right)^{-1}
%  \left[
  \Gamma_{\text{f}} + \dot{\varepsilon}(\Gamma_{\sigma,11}-\Gamma_{\sigma,22})
\quad\text{and}\quad
\overline{\Gamma}_{2} = 0
% \right]
\end{gathered}\label{eq:Gamma3}
\end{equation}
in which we have eliminated the leading term
$M^{\text{int}}\overline{\ell}^{2}p_{\text{o}}/A$ that appears in
Eqs.~(\ref{eq:GammaSigma}) and~(\ref{eq:Gammaf}).
By substituting Eqs.~(\ref{eq:N}), (\ref{eq:tc}),
(\ref{eq:GammaSigma}), and (\ref{eq:Gammaf}),
the function $\Gamma_{2}(\cdots )$ can be expanded as
\begin{align}
\Gamma_{2}(\cdots ) =
K_{2}( \theta^{\text{c}},\theta^{\ell})\,
&\left[
\sqrt{3/2}\,\dot{\phi}_{\text{slip}}\,g^{\text{t}}
\right.
%g^{\text{t}} (\dot{n} - \dot{\phi})\right. 
\label{eq:Gamma3Expand}\\
&\;+
\dot{\varepsilon}g^{\text{n}}
  (-\cos\theta^{\ell}\cos\theta^{\text{c}}
   +\sin\theta^{\ell}\sin\theta^{\text{c}})\notag\\
&\left.\; -\,
K_{1}(\theta^{\text{c}})\,
\dot{\varepsilon}\,
g^{\text{t}}\,
% g^{\text{t}}
  (\cos\theta^{\ell}\sin\theta^{\text{c}}
   +\sin\theta^{\ell}\cos\theta^{\text{c}})
\right] \notag
\end{align}
which enforces consistency of work and dissipation.
% It is also a virtual work statement of the global consistency
% of the internal stress and the internal contact forces,
% provided that the virtual displacements are restricted to those
% producing only tangential movements 
% %of form $\overline{\ell}(\dot{n}-\dot{\phi})$ 
% at the contacts \citep{Chang:2005a}.
% Although rate $\dot{\varepsilon}$ 
% appears explicitly in this equation,
% the rate will be treated later as simply
% a scaling parameter for normalizing
% $\dot{\phi}_{\text{slip}}$ and $\dot{\phi}_{\text{rigid}}$.
% %$\dot{n}$ and $\dot{\phi}$.
%
\subsection{Fundamental constraint~3: Isochoric flow} \label{sec:isochoric}
The third fundamental constraint (and, by far, the most difficult to derive) 
enforces the constant density (isochoric) condition of
critical state flow.
If tensor $L_{ij} = \partial v_{i}/\partial x_{j}$ is the average, bulk
velocity gradient within a granular assembly,
its trace vanishes during steady state flow.
To enforce this condition in an assembly
with $M$ contacts, we require a means of
estimating $\mathbf{L}$ from the contact motions of particle pairs: 
from an $M$-list of contact information
$\{ g^{\text{n}}_{k},g^{\text{t}}_{k},\theta^{\text{c}}_{k},\theta^{\ell}_{k},
\dot{\phi}_{\text{slip},k},\dot{\phi}_{\text{rigid},k} \}$.
Several methods have been proposed for estimating
$\mathbf{L}$ (see \citealp{Bagi:2006a} for a review), but most
require additional data that is not supplied in the $M$-list
of the six quantities.
Briefly, some methods involve a best-fit of particle
motions to an affine velocity field, % \citep[e.g.,][]{OSullivan:2003b},
but these methods require the particle locations and velocities.
Other methods use movements of boundary particles, but these
methods also require knowledge of the particles' locations as well
as their velocities.
Yet other methods use contact data alone, but require knowledge
of the topological adjacencies within the contact network.
%\citep{Bagi:1996a,Kruyt:1996a,Kuhn:1999a}.
%\citep{Bagi:1996a}.
\citet{Liao:1997a} 
have proposed an approximation of
$\mathbf{L}$ that requires only local contact data.
For the current work, we use an alternative approach,
which also requires \emph{only contact data} (rather than particle
data) but is
suitable for integration in the form of Eq.~(\ref{eq:GammaInt}).
This estimate
is derived from an analysis of the relative motions of contacting
particle pairs
that lie along the perimeter of a two-dimensional
granular assembly.
\par
\begin{figure}
\centering
\includegraphics{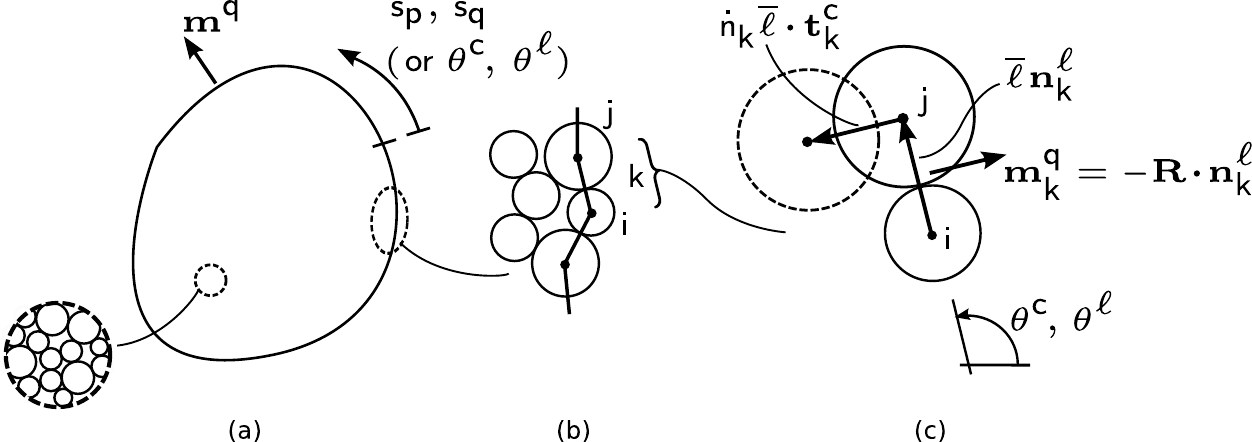}
\caption{Boundary chain of particles for finding the velocity gradient:
         (a)~bounded assembly of particles,
         (b)~chain of boundary branch vectors,
         and (c)~two boundary particles.
         \label{fig:deform}}
\end{figure}

The author has shown that the average velocity gradient
$\mathbf{L}$ within a two-di\-men\-sion\-al region
is exactly given by a double integral 
around its perimeter, in which perimeter locations are 
parameterized by the arc
distances $s^{p}$ and $s^{q}$ \citep{Kuhn:2004a}.  
Both distances are measured
counterclockwise from a common, fixed point on the 
boundary (Fig.~\ref{fig:deform}a).
The area-average velocity gradient is
\begin{align}
%\begin{aligned}
\mathbf{L} &= \frac{1}{A} \iint\limits_A
              \frac{\partial v_{i}}{\partial x_{j}}\,dA\\
 &= \frac{1}{A} \int_{0}^{S} \int_{0}^{S}
 \frac{dv_{i}}{ds^{p}}\, Q\left( s^{p}, s^{q}\right)
 m_{j}^{q} \, ds^{p} ds^{q}\label{eq:IntegralL}
%\begin{aligned}
\end{align}
where $S$ is the perimeter length;
$\mathbf{m}^{q}$ is the unit outward normal of the perimeter
at position $s^{q}$;
$A$ is the enclosed area;
$d v_{i}/d s^{p}$ is the derivative of the velocity component $v_{i}$ 
along the perimeter, as
point $s^{p}$ traverses this boundary;
and kernel $Q$ is a discontinuous function 
with domain $[0,S)$ and range $(-1/2,1/2)$
given by
\begin{equation}
\displaystyle
Q\left( s^{p}, s^{q}\right) =
\begin{cases}
  \displaystyle
  \frac{1}{2} - \frac{1}{S}\operatorname{mod}\left(s^{q}-s^{p},S\right),& s^{q}\neq s^{p}\\
  0,& s^{q}=s^{p}
\end{cases}\label{eq:Q1}
\end{equation}
using the modulo $\text{mod}(\,)$ function.
Integral (\ref{eq:IntegralL}) is an objective rate
and requires only two types of information
around the boundary:
the boundary normal $\mathbf{m}^{q}$ and the \emph{relative motion}
$d v_{i}/d s^{p}$
along the perimeter, which
is related to the contact motions.
%$d v_{i}/d s^{p}$.
The enclosed area $A$ can also be expressed in a similar
manner (Section~\ref{sec:auxiliary}).
Equation~(\ref{eq:IntegralL}) is a general and exact expression
for the average spatial gradient within
a closed two-dimensional region (bounded by a Jordan curve),
provided that the boundary derivative $dv_{i}/ds$ is integrable
and consistent with a field of boundary displacements,
such that $\int_{0}^{S}(dv_{i}/ds)\,ds = 0$.
\par
For an assembly of disks, we focus on the closed
polygonal chain of $M^{\text{per}}$ segments 
(branch vectors) formed by the
contacting disks along the assembly's perimeter (Fig.~\ref{fig:deform}b).
The perimeter segment $k$ between
particles $i$ and $j$ has outward unit normal $\mathbf{m}^{q}_{k}$ 
which is perpendicular
to the branch vector,
$\mathbf{m}^{q}_{k}=-\mathbf{R}\cdot\mathbf{n}^{\ell}_{k}$, with
arc length $\int_{i}^{j}ds^{q}=\overline{\ell}$
and rotation tensor $\mathbf{R}$
% In these and future expressions,
% $\mathbf{n}^{\text{c}}$ %(or $\mathbf{n}^{p}$)
% and $\mathbf{n}^{\ell}$ %(or $\mathbf{n}^{q}$) 
% are unit orientation vectors
% associated with the contact normal and branch vectors, and
% $\mathbf{R}$ is a rotation tensor
that effects 
a counterclockwise rotation of $\pi/2$:
\begin{equation}
\mathbf{R} = \begin{bmatrix} 0 & -1\\ 1& 0\end{bmatrix}\label{eq:R}
\end{equation}
\par
In applying Eq.~(\ref{eq:IntegralL}) to critical state flow,
we assume that the particles are rigid and that bulk deformation
is entirely due to the relative tangential motions between
contacting particles.
That is, we neglect any small changes in contact indentations
during deformation and neglect the opening (extinction) and
closing (creation) of contacts during small increments of bulk
deformation.
Such normal motions have a negligible contribution
to the bulk deformation
% This assumption is consistent with
% previous studies showing that
% movements in the contacts' normal directions
% (i.e., changing overlaps in DEM parlance) are
% small when compared with tangential movements during critical
% state, plastic flow 
(characteristic~A.1 in the Introduction).
%(Moreover, the critical state is also attained with simulations that
%model perfectly rigid disks, 
%altogether precluding contact indentations.)
%With this assumption,
Considering only tangential motions,
the velocity of particle $j$ relative to $i$ is
equal to the angular rate $\dot{n}_{k}$ multiplied by
branch length
$\overline{\ell}$ and by the unit tangential direction
$\mathbf{t}^{\text{c}}_{k}$, producing the rate vector
$\dot{n}_{k}\overline{\ell}\,\mathbf{t}^{\text{c}}_{k}$
(Fig.~\ref{fig:deform}c).
This \emph{relative velocity} occurs along a branch vector of length
$\overline{\ell}$, so that derivative 
$dv_{i}/ds^{p}=\dot{n}_{k}\mathbf{t}^{\text{c}}_{k}$.
Rate $\dot{n}_{k}$ can be expressed in terms of 
%$\dot{\phi}_{\text{slip},k}$ and $\dot{\phi}_{\text{rigid},k}$
``$\phi$'' rates
(see Eq.~\ref{eq:PhiRates}), as
\begin{equation}
\dot{n}_{k} = \left. \left(2\dot{\phi}_{\text{slip},k} 
  + \sqrt{2}\,\dot{\phi}_{\text{rigid},k}\right) \right/ \sqrt{6}
\label{eq:nphi}
%\sqrt{\frac{2}{3}}\dot{\phi}_{\text{slip},k}
%+ \sqrt{\frac{1}{3}}\dot{\phi}_{\text{rigid},k}
\end{equation}
The double integration in (\ref{eq:IntegralL}) distinguishes
between ``$p$'' and ``$q$'' quantities:
$s^{p}$ is associated with contact movement, whereas $s^{q}$ is associated
with the outer normal $\mathbf{m}^{q}$.
In a similar manner, we distinguish between the ``$\text{c}$'' and
``$\ell$'' directions, 
$\mathbf{t}^{\text{c}}_{k}$ and $\mathbf{n}^{\ell}_{k}$,
and write (\ref{eq:IntegralL}) as the double sum:
\begin{equation}
\mathbf{L} =
- \frac{\overline{\ell}^{2}}{A}
\sum_{p=1}^{M^{\text{per}}}
\sum_{q=1}^{M^{\text{per}}}
\dot{n}_{p}\,
Q\left(s^{p},s^{q}\right) \,
\mathbf{t}^{\text{c}}_{p} \otimes
\left( \mathbf{R}\cdot\mathbf{n}^{\ell}_{q} \right) \label{eq:LSum}
\end{equation}
where $M^{\text{per}}$ is the number of perimeter contacts, and
$s^{p}$ and $s^{q}$ are measured counterclockwise from a single, common
perimeter point.  
This double sum is an exact expression for average
strain in a mono-disperse assembly of disks, 
provided that contact movements are limited to tangential
displacement. 
% (Other terms are easily added to account for changes
% in the normal indentations, for non-circular shapes, 
% and for poly-dispersity,
% although such refinements are unnecessary in the current work.)
\par
To develop a corresponding
function $\Gamma(\cdots)$ that can be integrated as in
Eq.~(\ref{eq:GammaInt}),
we idealize the material region 
in the limit of a very large assembly of disks in which
the perimeter chain of contacts has an orientation 
$\theta$ that increases
monotonically from one contact to the next, forming
a convex, nearly smooth boundary.
In making this idealization, % of the perimeter chain,
we neglect the non-convex ``inside corners'' that would typically
occur around a polygonal loop of branch vectors.
Arc distances $s^{p}$ and $s^{q}$ 
can now be approximately
parameterized with angles $\theta^{\text{c}}$ 
and $\theta^{\ell}$.
We also assume that the 
probability density of these angles is the same
as that within the interior of the region:
the density $p(\cdot,\cdot,\theta^{\text{c}},\theta^{\ell},\cdot,\cdot)$.
The shape of the boundary chain will depend upon this density,
which furnishes the relative
numbers of contacts at different orientations.
With these assumptions,
Eqs.~(\ref{eq:IntegralL}), (\ref{eq:Q1}), and~(\ref{eq:LSum}) 
are written as
\begin{equation}
\mathbf{L} \approx 
\underset{( g^{\text{n}},g^{\text{t}}, 
        \theta^{\text{c}},\theta^{\ell},
        \dot{\phi}_{\text{slip}},\dot{\phi}_{\text{rigid}} )}
        {\idotsint} 
\boldsymbol{\Gamma}_{\mathbf{L}}(\cdots)\, p(\cdots ) \label{eq:IntegerL2}
\end{equation}
with function $\boldsymbol{\Gamma}_{\mathbf{L}}$
and kernel $K_{3}(\theta^{\text{c}},\theta^{\ell})$,
\begin{align}
\boldsymbol{\Gamma}_{\mathbf{L}}( \cdots )
&=
-\frac{\left(M^{\text{per}} \overline{\ell}\right)^{2}}{A}\,
K_{3}( \theta^{\text{c}},\theta^{\ell})
\,\dot{n}\,\mathbf{t}^{\text{c}}\otimes
(\mathbf{R}\cdot\mathbf{n}^{\ell})  \label{eq:GammaL} \\
K_{3}( \theta^{\text{c}},\theta^{\ell})
&=
\begin{cases}\displaystyle
\frac{1}{2} - \frac{1}{2\pi} \operatorname{mod} (\theta^{\ell}-\theta^{\text{c}},2\pi),
& \theta^{\ell}\neq\theta^{\text{c}}\\
0, & \theta^{\ell}=\theta^{\text{c}}
\end{cases}\label{eq:K2}
\end{align}
and with $\mathbf{n}^{\ell}$ and $\mathbf{R}$ defined
in Eq.~(\ref{eq:R}) and $\dot{n}$ defined in
Eq.~(\ref{eq:nphi}).
Unlike the Dirac kernel ($K_{2}$ in Eq.~\ref{eq:K1}),
the kernel $K_{3}$
of Eq.~(\ref{eq:K2}) effects the \emph{double sum}
of Eq.~(\ref{eq:LSum}), in which
the two arguments
($\theta^{\text{c}}$ and $\theta^{\ell}$, or
$s^{p}$ and $s^{q}$)
correspond to different points on the assembly boundary.
\par
Although Eqs.~(\ref{eq:IntegralL}) and~(\ref{eq:LSum})
yield exact average gradients,
Eqs.~(\ref{eq:IntegerL2})--(\ref{eq:K2}) are an estimate.
The author's DEM simulations show errors of 5\%--25\% 
for disk assemblies in critical state flow.
%Although the error can be substantial, the
Notwithstanding the small error, the
approximation (\ref{eq:IntegerL2}) is only intended 
as a modest constraint on
%to modestly constrain
the contact motions, holding them to a nearly isochoric condition.
Errors result from a number of sources:
(1) the non-tangential (normal) contact movements that can
occur between non-rigid grains;
(2) the approximation of $ds$ in Eq.~(\ref{eq:IntegralL}) with
$M^{\text{per}}\overline{\ell}p(\cdots)d\theta$ in Eq.~(\ref{eq:IntegerL2}),
which is a truncation of the full expansion of $s(\theta)$;
(3) the non-monotonic variation of $\theta$ around a non-convex perimeter
chain of branch vectors; and
(4) subtle correlations among the branch vector lengths $\ell$,
contact movements $\dot{n}$, and orientations 
$\theta^{\text{c}}$ and $\theta^{\ell}$.
\par
% Applying the estimated velocity gradient
% of Eqs.~(\ref{eq:IntegerL2})--(\ref{eq:tc}),
The isochoric constraint on the probability density $p(\cdots )$
is the requirement of a zero 
expected value of trace $\text{tr}(\mathbf{L})$,
expressed as
\begin{equation}
\begin{gathered}
\left\langle\Gamma_{3}(\cdots)\right\rangle = \overline{\Gamma}_{3}
\;\text{, where}
\\
\Gamma_{3}(\cdots) =
  K_{3}( \theta^{\text{c}},\theta^{\ell})
  \operatorname{tr}\left[ (\dot{n}\mathbf{t}^{\text{c}} )\otimes
  (\mathbf{R}\cdot\mathbf{n}^{\ell})\right]
  \quad\text{and}\quad
  \overline{\Gamma}_{3} = 0
\end{gathered}\label{eq:Gamma1}
\end{equation}
in which the leading constant term in (\ref{eq:GammaL}) is canceled.
The dyad in Eq.~(\ref{eq:Gamma1}) is
\begin{equation}
  \dot{n}\,\mathbf{t}^{\text{c}} \otimes
  (\mathbf{R}\cdot\mathbf{n}^{\ell}) =
  K_{1}(\theta^{\text{c}})\,
\dot{n}
\begin{bmatrix}
\sin\theta^{\text{c}} \sin\theta^{\ell} & -\sin\theta^{\text{c}} \cos\theta^{\ell}\\
-\cos\theta^{\text{c}} \sin\theta^{\ell} & \cos\theta^{\text{c}} \cos\theta^{\ell}
\end{bmatrix}
\end{equation}
with trace
\begin{equation}
  \operatorname{tr}\left[\dot{n}\mathbf{t}^{\text{c}}\otimes
  (\mathbf{R}\cdot\mathbf{n}^{\ell})\right] =
  K_{1}\,(\theta^{\text{c}})
  \dot{n}\left(\sin\theta^{\text{c}} \sin\theta^{\ell} + 
  \cos\theta^{\text{c}} \cos\theta^{\ell}\right)
\end{equation}
where $\dot{n}$ is given by Eq.~(\ref{eq:nphi}),
and function $K_{1}(\theta^{\text{c}})$ of Eq.~(\ref{eq:tc})
enforces the biaxial symmetry shown in
Fig.~\ref{fig:Conventions}.
\subsection{Fundamental constraint~4: Loading rate}\label{sec:auxiliary}
%
% Auxiliary moment constraints on $p(\cdots )$ can also
% be added, delivering other
% information that might be known \emph{a priori}.
% This information could include
% the stress or deformation (boundary) conditions, 
% the fraction of contacts that are sliding (Section~\ref{sec:eta}), 
% the average magnitude of the tangential contact forces, 
% the averaged nature of contact motion 
% (Section~\ref{sec:activity}), etc.
We must specify the deformation rate $\dot{\varepsilon}$ that was
% by establishing the scalar value $\dot{\varepsilon}$
% that applies to the specific symmetry of biaxial compression
% (Eq.~\ref{eq:Dmatrix}) and
% has already been
included % as a constant term
in the second (dissipation) constraint of
Eqs.~(\ref{eq:Gamma3}) and~(\ref{eq:Gamma3Expand}).
Applying Eqs.~(\ref{eq:IntegerL2}) and~(\ref{eq:GammaL}) 
to the single deformation
component $L_{11}=-\dot{\varepsilon}$
of Eq.~(\ref{eq:Dmatrix}) yields the following
constraint on probability density $p(\cdots )$:
\begin{equation}
-\frac{\left(M^{\text{per}} \overline{\ell}\right)^{2}}{A}
\underset{( g^{\text{n}},g^{\text{t}},
        \theta^{\text{c}},\theta^{\ell},
        \dot{\phi}_{\text{slip}},\dot{\phi}_{\text{rigid}} )}
        {\idotsint}
K_{3}( \theta^{\text{c}},\theta^{\ell})
\,\dot{n}\,
\left[ \mathbf{t}^{\text{c}}\otimes
       ( \mathbf{R}\cdot\mathbf{n}^{\ell})\right]_{,11}
%\sin\theta^{\text{c}} \sin\theta^{\ell}
=
-\dot{\varepsilon}  \label{eq:L11}
\end{equation}
We can eliminate
the leading coefficient $(M^{\text{per}} \overline{\ell})^{2}/A$
by again considering the
deformation gradient of Eqs.~(\ref{eq:IntegralL}), 
(\ref{eq:IntegerL2}),
and~(\ref{eq:GammaL}), but for the special case of an ideal
dilation, with $dv_{1}/ds^{p}=n_{1}^{p}$ and
$dv_{2}/ds^{p}=n_{2}^{p}$, which produces a known deformation rate with
trace $\text{tr}(\mathbf{L})=2$:
\begin{equation}
-\frac{\left(M^{\text{per}} \overline{\ell}\right)^{2}}{A}
\underset{( g^{\text{n}},g^{\text{t}}, 
        \theta^{\text{c}},\theta^{\ell},
        \dot{\phi}_{\text{slip}},\dot{\phi}_{\text{rigid}} )}
        {\idotsint}
K_{3}( \theta^{\text{c}},\theta^{\ell})
\,\text{tr}\left( \mathbf{n}^{\text{c}}\otimes
(\mathbf{R}\cdot\mathbf{n}^{\ell}) \right)
=
2 \label{eq:LDilate}
\end{equation}
Dividing Eq.~(\ref{eq:L11}) by Eq.~(\ref{eq:LDilate}) 
and rearranging terms gives the fourth constraint
on density $p(\cdots)$:
\begin{equation}
\begin{gathered}
\left\langle\Gamma_{4}(\cdots)\right\rangle = \overline{\Gamma}_{4}
\;\text{, where}\\
%  -\dot{\varepsilon} 
%  \left(\frac{\left(M^{\text{per}} 
%  \overline{\ell}\right)^{2}}{A}\right)^{-1} \\
\Gamma_{4}(\cdots ) = 
  K_{3}( \theta^{\text{c}},\theta^{\ell})
  \left\{
      \frac{\dot{\varepsilon}}{2}\,\text{tr}
      \left[ \mathbf{n}^{\text{c}}\otimes
       ( \mathbf{R}\cdot\mathbf{n}^{\ell})\right] +
      \dot{n}
      \left[ \mathbf{t}^{\text{c}}\otimes
       ( \mathbf{R}\cdot\mathbf{n}^{\ell})\right]_{,11}\right\}\\
  \quad\text{and}\quad
  \overline{\Gamma}_{4} = 0
%  \left( \frac{1}{2}K_{3}(\theta^{\text{c}})\dot{\varepsilon}
%  (  \sin\theta^{\text{c}} \cos\theta^{\ell} 
%   - \cos\theta^{\text{c}} \sin\theta^{\ell})
%  +
%  \dot{n}
%  \sin\theta^{\text{c}}\sin\theta^{\ell}
%  \right)
\end{gathered}\label{eq:Gamma4}
\end{equation}
which formally asserts the compression rate 
$\mathbf{L}_{11}=-\dot{\varepsilon}$,
assuring consistency of 
second and third constraints (Eqs.~\ref{eq:Gamma3}
and~\ref{eq:Gamma1}).
The function $\Gamma_{4}(\cdots)$ is computed as
\begin{equation}
\Gamma_{4}(\cdots) =
K_{3}( \theta^{\text{c}},\theta^{\ell})
  \left( \frac{\dot{\varepsilon}}{2}K_{1}(\theta^{\text{c}})
  (  \sin\theta^{\text{c}} \cos\theta^{\ell} 
   - \cos\theta^{\text{c}} \sin\theta^{\ell})
  +
  \dot{n}
  \sin\theta^{\text{c}}\sin\theta^{\ell}
  \right)
\end{equation}
with $\dot{n}$ given by Eq.~(\ref{eq:nphi}).
\par
\subsection{Auxiliary constraint 5:
            Fraction of sliding contacts}\label{sec:eta}
The four constraints given above describe a model of critical state flow
under biaxial loading.
Although a simple four-constraint model will predict nearly all
\emph{trends} observed in experiments and numerical simulations,
the qualitative agreement is, in some respects, in poorer accord:
in particular, it over-predicts the %deviatoric stress, fabric
% anisotropy, 
fraction of sliding contacts and the
activity of contact movements (a prediction of over 80\% sliding contacts).
These aspects can be improved with the
% Reasons for such results are discussed in the paper's conclusion,
% but we now consider three additional 
% constraints that greatly improve the quantitative predictions.
% Introducing these constraints, however,
% requires a
supply of new information to the model --- information 
of an entirely empirical origin.
We include a single modest parcel of information.
% This \emph{a priori} information would not normally be available
% from experiments, but can be readily extracted from simulations:
% in our case, from the DEM simulations that are described below.
% Two of the three constraints concern the
% vigor and form of contact movement.
A four-constraint model predicts an excess
in the fraction
of sliding contacts
(more than 80\%), compared with the
much smaller values noted in characteristic C.1 of the Introduction.
The final constraint limits the fraction
of sliding contacts $\eta$: %to that
%found in DEM simulations:
%
\begin{equation}
\begin{gathered}
\left\langle\Gamma_{5}(\cdots)\right\rangle = \overline{\Gamma}_{5}
\;\text{, where}
\\
\overline{\Gamma}_{5} = \eta
\;\;\text{and}
\\
\Gamma_{5}(\cdots) = 
  \begin{cases}
    1, & g^{\text{t}} \in \{-\mu g^{\text{n}},\mu g^{\text{n}}\}\\
    0, & g^{\text{t}} \in (-\mu g^{\text{n}},\mu g^{\text{n}})
  \end{cases}
\end{gathered} \label{eq:Gamma7}
\end{equation}
where a value of $\eta$ is empirically derived.
%For a friction coefficient $\mu=0.50$, the fraction of sliding 
%contacts $\eta$ was found to be 11.2\% in the simulations.
% \par
% This fifth constraint completes the model for dense granular
% flow, one that we consider a ``minimal model.''
% As will be seen, the model leads to excessively active
% contact motions, and a ``modified model'' will occasionally be
% referenced, in which an additional constraint is placed upon
%these motions.
%
%
\subsection{Entropy}\label{sec:Entropy}
In the view of an experimentalist or simulator,
a granular medium is a deterministic 
system --- although one of bewildering complexity --- in which
the evolving conditions of each contact (its motions, forces, etc.)
are uniquely determined by its own condition, 
%and those of its two particles, 
by the conditions of all other
contacts in the system, 
and by the boundary motions and forces.
The probabilities $p(\cdots)$ are accessible, in this view,
by frequent empirical observation of the contacts: from data 
in the form of
the $M^{6}$ ``$k$'' values.
\citet{Jaynes:1957a} describes this 
as an ``objective'' approach to
probability, when probability is an
expressed expectation based on observation.
To Jaynes,
statistical mechanics is based upon an
alternative ``subjective'' school
of thought in which probabilities are simply expressions of
expectation based upon general information that is usually very limited.
Any shortcomings of such predictions, when compared
with experimental observation, are attributed by the
``subjectivist'' not to insufficient acuity but
to insufficient information.
Indeed, to the subjectivist, 
the notion of probability is irrelevant when
all information of a deterministic system is available beforehand:
such an exercise is not one of prediction but of certitude.
\par
From a subjective viewpoint, 
this lack of information is synonymous with
uncertainty, disorder, or ``entropy'' and is quantified, 
in our case, with the differential Shannon entropy
\begin{equation}
H\left( p(\cdots)  \right) = \!
\underset{( g^{\text{n}},g^{\text{t}}, 
        \theta^{\text{c}},\theta^{\ell},
        \dot{\phi}_{\text{slip}},\dot{\phi}_{\text{rigid}} )}
        {\idotsint}\!\!\!\!
   p(\cdots) \ln\left( p(\cdots) \right)  \label{eq:H}
\end{equation}
where $p(\cdots)$ must satisfy the constraints of
the previous sections.
Although restrictive, this small set of
constraints does not, by itself, determine a unique macro-state, 
as many macro-states will satisfy the same constraints.
The most likely macro-state --- the most likely
probability density $p(\cdots)$ --- is the one
that encompasses
the greatest breadth of admissible micro-states among the
$M^{6}$ space of possibilities
%\citep[see][]{Shannon:1948a,Jaynes:1957a,BenNaim:2008a}.
\citep[see][]{Jaynes:1957a}.
For a large sample size (i.e. large $M$),
this probability density maximizes $H$ while satisfying any
available information that restricts the admissible
micro-states:
information in the form of our five moment constraints
\begin{equation}\label{eq:ConstraintsB}
\left\langle \Gamma_{i}(\cdots)\right\rangle = \overline{\Gamma}_{i}
,\quad i=1,2,\ldots,5
\end{equation}
%
%presented above. 
%(with the simplified model
%of Section~\ref{sec:ModelII}, $i=0 , \ldots , 3$).
If, instead, a granular system was to settle upon a
macro-state of \emph{lower} entropy,
this occurence would indicate a bias toward
greater order (a macro-state with a smaller breadth of micro-states),
suggesting an influence of other unaccounted information 
(i.e., other constraints).
This possibility is addressed in the conclusions.
\par
Following the Jaynes formalism of maximizing $H$ (i.e., the
``MaxEnt'' principle), the condition
\begin{equation}
\frac
{\partial H(p(\cdots))}
{\partial p(\cdots)}
=0\label{eq:MaxEnt}
\end{equation}
and the zeroth constraint of Eq.~(\ref{eq:Certainty}) lead to
the most likely density
\begin{equation}
p(\cdots) =
\frac{1}{Z(\cdots)}
\exp \left(
-\sum_{i=1}^{5}
\lambda_{i}\Gamma_{i}(\cdots)
\right)\label{eq:MaxP}
\end{equation}
with normalizing (partition) function $Z$,
\begin{equation}
Z(\cdots) = 
\underset{( g^{\text{n}},g^{\text{t}}, 
        \theta^{\text{c}},\theta^{\ell},
        \dot{\phi}_{\text{slip}},\dot{\phi}_{\text{rigid}} )}
        {\idotsint}
\exp \left(
-\sum_{i=1}^{5}
\lambda_{i}\Gamma_{i}(\cdots)
\right) \label{eq:MaxZ}
\end{equation}
and with five Lagrange multipliers $\lambda_{i}$ that are computed
so that the five moment constraints
%Eqs.~(\ref{eq:Gamma1}), (\ref{eq:Gamma2}), (\ref{eq:Gamma3}), 
%and~(\ref{eq:Gamma4}) 
are satisfied
%\citep[again,][]{Shannon:1948a,Jaynes:1957a,BenNaim:2008a}.
\citep[again,][]{Jaynes:1957a}.
These multipliers are computed as
the solutions of five non-linear equations,
requiring a rather taxing evaluation of multiple
multi-dimensional integrations,
as described in \ref{sec:Numerics}.
%Once found, the multipliers fully define $p(\cdots)$, from which
%more meaningful information can be extracted.
% Solving the multipliers is detailed in
% Section~\ref{sec:Implementation} and \ref{sec:Numerics}.
%
%
\section{DEM Simulations and Model Verification}\label{sec:Compare}
This section describes the small set of parameters that
are required to implement and solve
the model (Section~\ref{sec:Implementation}),
the DEM simulations that were used to evaluate the model
(Section~\ref{sec:DEM}),
and the results of this evaluation
(Section~\ref{sec:compare1}).
\subsection{Model implementation}\label{sec:Implementation}
The model requires three input parameters:
the inter-particle friction coefficient $\mu$,
the fraction of sliding contacts $\eta$,
and the contact density
coefficient $(M^{\text{int}}\overline{\ell}^{2}/A)^{-1}$.
Mean stress $p_{\text{o}}$ and strain rate $\dot{\varepsilon}$
also appear, but only as scaling factors, and the forces and sliding
rates are normalized with respect to these factors.
Coefficient $(M^{\text{int}}\overline{\ell}^{2}/A)^{-1}$ is
largely a scaling factor that does not significantly affect results
and was was assigned a value of 1.5 in all calculations %, a value
% determined with the DEM simulations
(see the end of Section~\ref{sec:mean}).
Results depend primarily on the friction coefficient, and
a range of values were used
($\mu=0.1$, 0.3, 0.5, 0.7, and 0.9),
with most results reported for $\mu=0.5$.
The fraction $\eta$ ranged from 5\% to 37\% in the simulations,
depending on
friction $\mu$.
An average value of 15\%
was used with the model.
\par
Equations~(\ref{eq:ConstraintsB})--(\ref{eq:MaxZ}) 
define the density $p(\cdots)$ that corresponds
to maximum disorder,
a density expressed in terms of five multipliers $\lambda_{i}$.
Once the $\lambda_{i}$ are computed (\ref{sec:Numerics}), 
we can extract meaningful information 
%of the idealized granular system
from the model
by evaluating various expected values and
marginal distributions of $p(\cdots )$.
These evaluations and an associated notational system are
described in \ref{app:Conventions}.
\subsection{DEM simulations}\label{sec:DEM}
\begin{figure}
\centering
\includegraphics{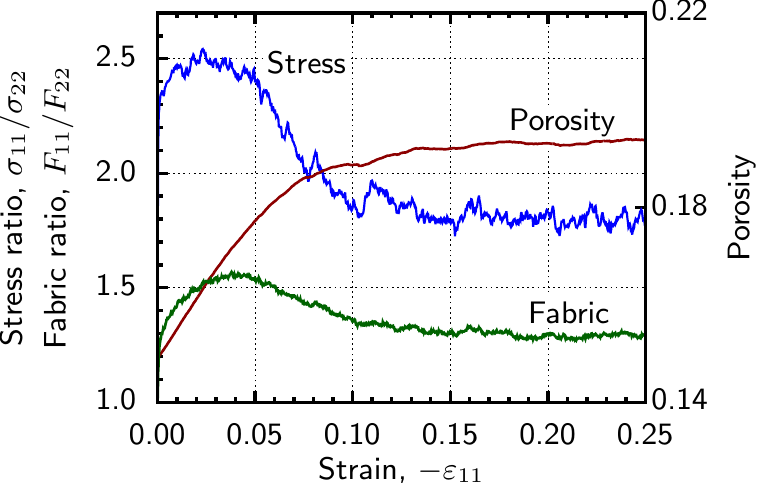}
\caption{Stress, fabric, and porosity from DEM simulations
         (friction coefficient $\mu=0.50$).}
\label{fig:crs}
\end{figure}

Statistics of contact force, movement, and orientation
were gathered from simulations of 
168 assemblies, each with 676 bi-disperse disks,
which were sheared in horizontal biaxial compression.
Details of the simulations are found in \ref{app:DEM}.
The bulk behavior is illustrated in Fig.~\ref{fig:crs}, which
shows the evolution of stress,
% $\sigma_{11}/\sigma_{22}$,
fabric, and porosity.
%the fabric tensor ratio $F_{11}/F_{22}$, and the porosity.
The Satake fabric tensor $F_{ij}$ is the
average $\langle n_{i}^{\text{c}}n_{j}^{\text{c}}\rangle$
of the contact orientation vectors $\mathbf{n}^{\text{c}}$,
and the ratio $F_{11}/F_{22}$
is a fundamental measure of bulk anisotropy
% (note, the trace of $\mathbf{F}$ is $1.0$)
\citep{Satake:1982a}.
A large initial stiffness causes the 
stress ratio $\sigma_{11}/\sigma_{22}$
to rise quickly from 1.0 to a peak condition at
strain $-\varepsilon_{11}=2\%$,
and the critical state condition is
reached at compressive strains 
%$-\varepsilon_{11}$ 
beyond 16--18\%,
with a mean value of ratio $\sigma_{11}/\sigma_{22}$ of
1.80. % and standard deviation XX.
While in the critical state, the bulk response is seen
to fluctuate (with a standard deviation of 0.03), 
indicating the agitated, tumultuous nature
of the underlying particle interactions.
%The critical state was reached at 
%strain $-\varepsilon_{11}$ of 16--18\%.
\par
By running 168 simulations
with random initial configurations and by taking several snapshots
during the subsequent critical state flow, we collected over 800,000
contact samples of orientations, forces, and movements.
Once the critical state is reached, we found that any bulk
average of the micro-scale data
(stress, fabric, density, etc.) or any marginal distribution thereof
attains a nearly constant, steady condition.
%
%\subsection{Conventions}\label{sec:Conventions}
%To compare results of the two entropy models with
%simulation results,
%we begin by establishing two conventions:
%a distinction between sliding in forward and reverse directions,
%and notations for expressing certain statistical quantities.
%Figure~XX shows the directions of sliding and tangential force that 
%would be expected across the surface of particle $i$
%during horizontal biaxial compression, provided that
%particle movements roughly conform to an affine displacement field.
%Although such ``forward'' sliding is common, we will see that
%many contacts undergo ``reverse'' sliding.
%A kernel $K_{3}(\dot{n}-\dot{\phi},\theta^{\ell})$ is used to specify
%sliding direction within integrations, such as Eq.~(\ref{eq:GammaInt}),
%
%\begin{equation}
%K_{3}(\dot{n}-\dot{\phi},\theta^{\ell})=
%\begin{cases}
% -1& n_{1}^{\ell}n_{2}^{\ell}(\dot{n}-\dot{\phi})>0\\
% 0&  (\dot{n}-\dot{\phi})=0\\
%1& n_{1}^{\ell}n_{2}^{\ell}(\dot{n}-\dot{\phi})<0
%\end{cases}
%\end{equation}
%
%where values $-1$ and 1 mean reverse and forward
%sliding, 0 means no sliding, and $n_{1}^{\ell}=\cos\theta^{\ell}$
%and $n_{2}^{\ell}=\sin\theta^{\ell}$.
%
%
%
%In doing so, we must compute various conditional probabilities,
%conditional expectations, and conditional probability densities.
%
%
\subsection{Comparison of model and DEM results}\label{sec:compare1}
The Introduction recounts numerous micro-scale characteristics
of granular flow.
We will find strong \emph{qualitative} agreement with
the DEM results:
with most characteristics described in the Introduction,
the model and simulations exhibit the same trends.
In presenting the model results,
the following paragraphs are preceded with raised items
(e.g., ``B.7'') that refer to particular characteristics
in the Introduction, and the
corresponding references to the literature can be 
found there.
Primary predictions of the entropy model are shown in
Table~\ref{table:Results1}, with references to the 
Introduction shown as
raised items in the first column.
The third column in Table~\ref{table:Results1} defines
the various quantities, using the notation of
\ref{app:Conventions}.
Only averages are reported, although dispersions from the
mean can also be extracted with the model.
Unless otherwise noted,
the results of the model and the simulations
are %now reported in a dimensionless and
% normalized form
for the single case $\mu=0.50$.
\par
In the following,
the normal contact forces $g^{\text{n}}$
are divided by
$2(M^{\text{int}}\overline{\ell}^{2}/A)^{-1}$
(i.e., the $\overline{\Gamma}_{2}$ factor in 
Eq.~\ref{eq:Gamma2}), so that the mean normed value
$\langle g^{\text{n}}\rangle_{\text{normed}}$ is 1
(row 6 of Table~\ref{table:Results1}).
In terms of an actual force $f^{\text{n}}$,
the normalized
$g^{\text{n}}_{\text{normed}}$ equals
$f^{\text{n}}(M^{\text{int}}\overline{\ell}/2p_{\text{o}}A)$.
% Movements $\dot{\phi}_{\text{slip}}$ and 
% $\dot{\phi}_{\text{rigid}}$ are normalized
% by dividing by strain rate $\dot{\varepsilon}$.
%
%
\begin{table}
\centering\small
\caption{Results of DEM simulations and model
         (friction coefficient $\mu=0.50$).
         Raised items in the first column refer 
         to characteristics in the Introduction.
         \label{table:Results1}}
\begin{tabular}{lp{2.00in}lcc}
\toprule
%heading 1
&&   & \multicolumn{2}{c}{Values}\\
\cmidrule{4-5}
%heading2
%&&&& Complete & Simple\\
%heading3
 & Description & \multirow{-2}{*}{\minitab[l]{Definition\\(see \ref{app:Conventions})}} & DEM & Model\\
\midrule
1$^{\text{(B.1)}}$ & 
    Fabric tensor ratio, $F_{11}/F_{22}$ & 
    $\langle n_{1}n_{1}\rangle / \langle n_{2}n_{2}\rangle$ &
    1.30 & 1.42\\
2$^{\text{(B.2)}}$ & 
    Deviatoric stress ratio, $q/p_{\text{o}}$ &
    $\langle \Gamma_{\sigma,11} - \Gamma_{\sigma,22} \rangle 
       / p_{\text{o}}$ &
    0.573 & 0.605\\
3$^{\text{(B.3)}}$ & 
    Tangential force contribution to the deviatoric stress ratio, 
      $q^{\text{t}}/p_{\text{o}}$ &
    $^{*\,}$$\alpha \langle g^{\text{t}}(n_{1}t_{1} - n_{2}t_{2}) \rangle
      /p_{\text{o}}$ &
    0.068 & 0.045\\
4$^{\text{(B.3)}}$ & 
    Normal force contribution to the deviatoric stress ratio,  
      $q^{\text{n}}/p_{\text{o}}$ &
    $^{*\,}$$\alpha \langle -g^{\text{n}}(n_{1}n_{1} - n_{2}n_{2}) \rangle
      /p_{\text{o}}$ &
    0.505 & 0.560\\
5$^{\text{(B.3)}}$ & 
    Percent contribution of tangential forces to deviatoric stress &
    $q^{\text{t}}/q$ &
    11.9\% & 7.5\%\\
6& Mean normal contact force, normalized &
    $\langle g^{\text{n}}\rangle_{\text{normed}}$ &
    1 & 1\\
7$^{\text{(B.4)}}$ & Fraction of deviator stress from weak contacts & 
    $\langle \Gamma_{\sigma,11}-\Gamma_{\sigma,22}\rangle
      |_{g^{\text{n}}<1}$ & 
    7.2\% & 11.0\%\\
8$^{\text{(B.4)}}$ & Fraction of mean stress from weak contacts & 
    $\frac{1}{2}\langle \Gamma_{\sigma,11}+\Gamma_{\sigma,22}\rangle
      |_{g^{\text{n}}<1}$ & 
    28.7\% & 26.7\%\\
9$^{\text{(B.5)}}$ & Fabric tensor ratio among weak contacts & 
    $  \langle n_{1}n_{1}\rangle|_{g^{\text{n}}<1} 
     / \langle n_{2}n_{2}\rangle|_{g^{\text{n}}<1}$ & 
    1.05 & 1.17\\
10$^{\text{(B.5)}}$ & Fabric tensor ratio among strong contacts & 
    $  \langle n_{1}n_{1}\rangle|_{g^{\text{n}}>1} 
     / \langle n_{2}n_{2}\rangle|_{g^{\text{n}}>1}$ & 
    1.76 & 1.80\\
11$^{\text{(C.1)}}$& Fraction of sliding contacts, $\eta$ &
    $\langle 1\rangle |_%
      {g^{\text{t}}\in\{-\mu g^{\text{n}},\mu g^{\text{n}}\}}$ &
    11.2\% & 15.0\%\\
12$^{\text{(B.6)}}$ & Fraction of forward sliding contacts, $\eta^{+}$ &
    $\langle 1\rangle |_%
      {g^{\text{t}}=\mu g^{\text{n}}}$ &
    6.6\% & 7.7\%\\
13$^{\text{(B.6)}}$ & Fraction of reverse sliding contacts, $\eta^{-}$ &
    $\langle 1\rangle |_%
      {g^{\text{t}}=-\mu g^{\text{n}}}$ &
    4.6\% & 7.3\%\\
14$^{\text{(B.6)}}$ & Ratio of forward and reverse sliding contacts &
    $\eta^{+}/\eta^{-}$ &
    1.43 & 1.05\\
15$^{\text{(B.8)}}$ & Mean normal contact force among non-sliding contacts &
    $\langle g^{\text{n}} | g^{\text{t}}\in(-\mu g^{\text{n}},\mu g^{\text{n}})\rangle_{\text{normed}}$ &
    1.073 & 1.15\\
16$^{\text{(B.8)}}$ & Mean normal contact force among forward sliding contacts &
    $\langle g^{\text{n}} | g^{\text{t}}=\mu g^{\text{n}}\rangle_{\text{normed}}$ &
    0.335 & 0.174\\
17$^{\text{(B.8)}}$ & Mean normal contact force among reverse sliding contacts &
    $\langle g^{\text{n}} | g^{\text{t}}=-\mu g^{\text{n}}\rangle_{\text{normed}}$ &
    0.306 & 0.159\\
\midrule
\multicolumn{5}{l}{$^{*}$ $\alpha=M^{\text{int}}\overline{\ell}^{2}p_{\text{o}}/A$}\\
\bottomrule
\end{tabular}
\end{table}%
%
%Both the complete and simplified models correctly predict that a minority 
%of the contacts undergo frictional slip:
%33\% for the complete model, and 11.2\% for the simplified
%model.
\par
\begin{figure}
\centering
\includegraphics{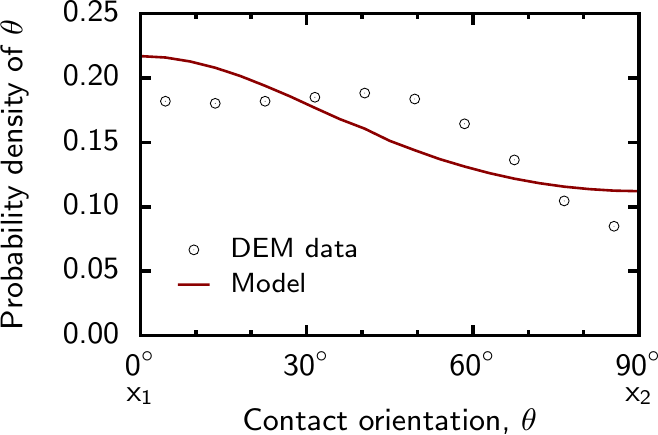}
\caption{Density distribution of contact orientation 
         for horizontal
         biaxial compression in the $x_{1}$ direction,
         $\mathcal{D}_{\theta^{\text{c}}} \langle 1\rangle$.
         Orientation $\theta=0$ corresponds to contact normals
         that are aligned with the compression direction
         ($x_{1}$ in Fig.~\ref{fig:Conventions}).
         \label{fig:results5}}
\end{figure}

$^{(\text{B.1})}$%
As the most telling result,
the model predicts an anisotropy of contact orientation
that is consistent with the simulations: contacts are predominantly
oriented in the direction of the major principal compressive
stress, with
a Satake fabric ratio 
$F_{11}/F_{22}=\langle n_{1}n_{1}\rangle/\langle n_{2}n_{2}\rangle$
that is greater than one
(row~1 of Table~\ref{table:Results1}).
The distribution of contact orientations
is shown in Fig.~\ref{fig:results5}, 
in which the $[0,2\pi)$ range has been folded 
to $[0,\pi/2)$ in
accordance with the biaxial symmetry of loading.
In this and other figures, the $x_{1}$ direction (i.e., $\theta=0$)
refers to contact normals that are oriented in the
(horizontal) direction of compressive loading,
and $x_{2}$ is in the direction of extension
(Fig.~\ref{fig:Conventions}).
Although the model and simulations display some differences,
the model correctly predicts the general trend of the 
orientation distribution.
\par
\begin{figure}
\centering
\mbox{%
%  \subfigure[]{\includegraphics%
%  {../Figures/Plot_7_paper_Circles-bidisperse-676-a_0500.pdf}}
%  \ \quad
%  \subfigure[]{\includegraphics%
%  {../Figures/Plot_8_paper_Circles-bidisperse-676-a_0500.pdf}}}
\includegraphics{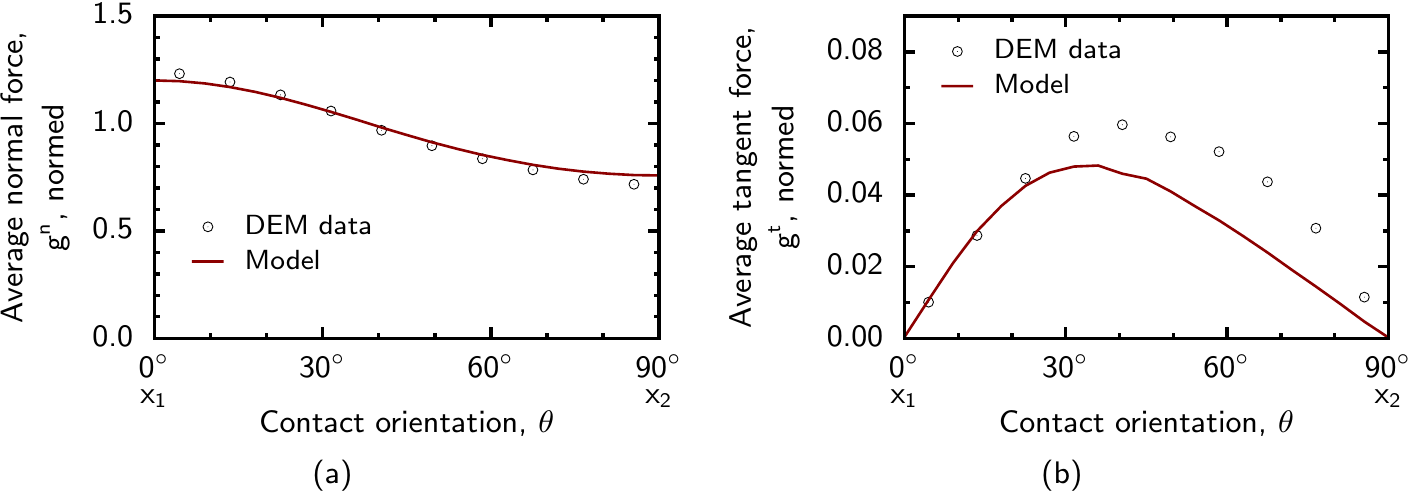}}
\caption{Average contact force as a function of contact orientation:
         (a) average normal contact force,
         $\langle g^{\text{n}} \rangle (\theta^{\text{c}})$,
         and (b) average tangential contact force,
         $\langle g^{\text{t}} \rangle (\theta^{\text{c}})$.
         \label{fig:results7and8}}
\end{figure}

$^{(\text{B.2})}$%
The model predicts that stress in the compression direction,
$-\sigma_{11}$, is larger than that in the extension
direction, $-\sigma_{22}$,
and the model predicts a deviator stress ratio,
$q/p_{\text{o}}=-(\sigma_{11}-\sigma_{22})/p_{\text{o}}$
that is close to the value from the DEM simulations 
(Table~\ref{table:Results1}, row~2).
% Although this agreement
% will be seen as somewhat coincidental,
% the model does predict an anisotropic stress with
% a ratio $\sigma_{11}/\sigma_{22}$ that is greater than 1.
As in Eqs.~(\ref{eq:SumStress}), (\ref{eq:IntSigma}),
and~(\ref{eq:GammaSigma}),
deviatoric stress can result from 
three sources: from contact orientations biased in
the compression direction, 
from larger normal contact forces in this same direction,
and from the tangential contact forces.
The predicted anisotropy of contact orientation was verified above
(Fig.~\ref{fig:results5} and row~1).
% Although a mean, normalized value of 1 was imposed on
% $g^{\text{n}}$,
The model also predicts an
anisotropy of the average
normal forces, as is illustrated in Fig.~\ref{fig:results7and8}a.
The results of the simulations and model are similar: normal forces
are, on average, larger in the direction of bulk compression.
Anisotropy of tangential contact force is shown in
Fig.~\ref{fig:results7and8}b.
For both simulations and model, the averaged tangential
force is largest at an angle of about 40$^{\circ}$
oblique to the compression direction.
Although the numerical values of simulations and theory differ,
the model captures the primary trend of anisotropy in the
tangential force.
\par
$^{(\text{B.3})}$%
The separate contributions of the normal and tangential forces to
the full deviatoric stress can be extracted with the model,
as these contributions are proportional
to the expected values 
$\langle -g^{\text{n}}(n_{1}n_{1} - n_{2}n_{2}) \rangle$
and
$\langle g^{\text{t}}(n_{1}t_{1} - n_{2}t_{2}) \rangle$.
For both simulations and model,
the relative contribution of the tangential forces to the 
full deviator stress $q$ is minor: a 11.9\% 
contribution in the simulations
and 8.9\% from the model (rows 3--5, Table~\ref{table:Results1}).
\par
\begin{figure}
\centering
\mbox{%
%  \subfigure[]{%
%  \includegraphics%
%  {../Figures/Plot_9_paper_Circles-bidisperse-676-a_0500.pdf}}
%  \ \quad
%  \subfigure[]{\includegraphics%
%  {../Figures/Plot_11_paper_Circles-bidisperse-676-a_0500.pdf}}}
   \includegraphics{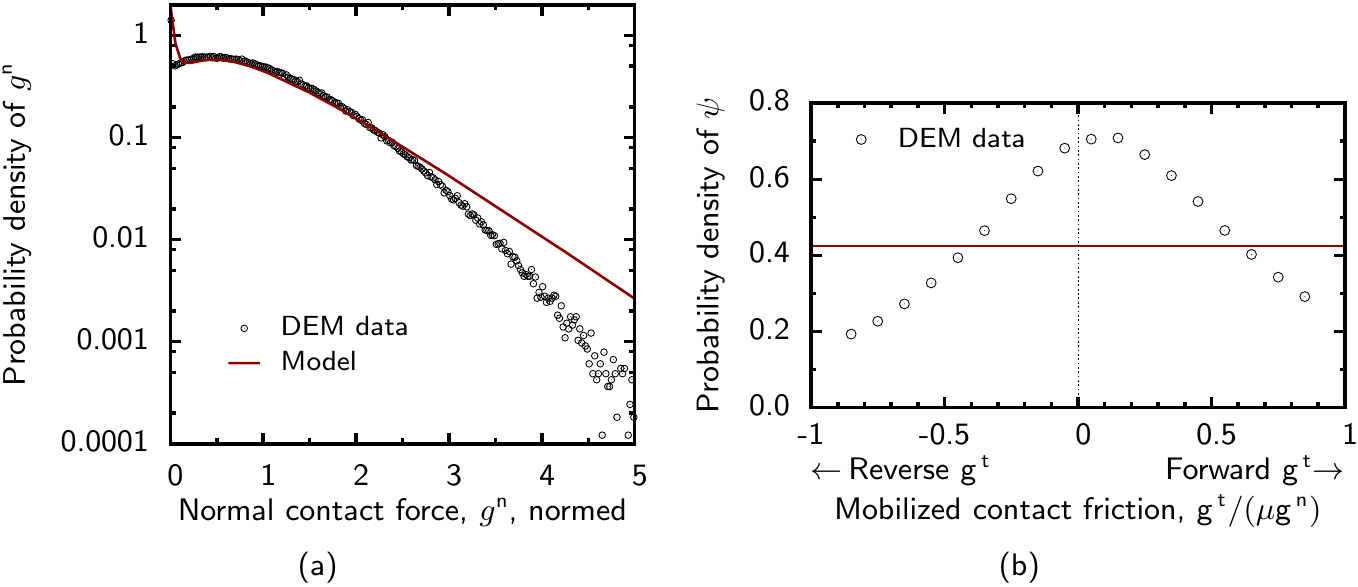}}
\caption{Density distributions of contact force:
         (a) normal force distribution,
         $\mathcal{D}_{g^{\text{n}}} \langle 1\rangle$,
         and (b) distribution of the
         fraction of mobilized friction of tangential forces,
         $\mathcal{D}_{g^{\text{t}}/\mu g^{\text{n}}} \langle 1\rangle$.
         \label{fig:results9and11}}
\end{figure}

$^{(\text{B.11})}$%
The model gives a fairly good prediction of the probability
density of normal contact forces, as seen in 
Fig.~\ref{fig:results9and11}a.
The model predicts an exponential tail for large normal
forces $g^{\text{n}}$ (although somewhat less steep
than the DEM data),
and it predicts a curved, flattened ``shoulder''for normal
forces $g^{\text{n}}$ less than 1.5.
The model, however, does give a steep rise in the distribution for
the smallest normal forces (those with $g^{\text{n}}<0.1$).
% The latter nuance is entirely removed with a modification to the model,
% by adding a constraint on the average magnitude of the
% slip motions $|\dot{\phi}_{\text{slip}}|$.
%
\par
$^{(\text{C.2})}$%
% it is unable to capture the flattened shoulder for normal
% forces $g^{\text{n}}$ less than the mean
% (equal to 1 in the normalized results), as has been consistently
% reported in the literature.
The model poorly predicts the
density of the tangential contact forces %, expressed as a fraction
% of mobilized friction is rather poor
(see the plot of $g^{\text{t}}/\mu g^{\text{n}}$
in Fig.~\ref{fig:results9and11}b).
When considering only the non-sliding contacts
(with $|g^{\text{t}}|<\mu g^{\text{n}}$),
the density of this ratio in the DEM simulations
decreases with an increasing mobilized friction,
consistent with experimental results \citep{Majmudar2007a}.
The model, however, gives a nearly uniform density for this same
condition.
This shortcoming is discussed in the conclusions.
\par
$^{(\text{B.4,B.5})}$%
Contacts with a normal force less than the mean force (weak contacts)
are known to operate differently than those with a 
greater-than-mean normal force (strong contacts).
The orientation distribution of weak contacts is
known to be nearly isotropic, with
only a small bias in the direction of loading.
Consistent with the simulations, the model predicts
% For both simulations and model,
a Satake fabric ratio for weak contacts of only slightly greater than one
(Table~\ref{table:Results1}, row~9); whereas the corresponding
value for the strong contacts is 1.8 (row~10):
that is, the contact anisotropy of all contacts (=1.42) is almost
entirely attributed to the strong contacts.
The deviatoric stress is also primarily attributed to the strong
contacts: with the simulations and the model, 11\% or less of the
deviator stress $q$ is derived from the weak contacts (row~7).
The same weak contacts have a more significant role in bearing
the mean stress, as more than 26\% of $p_{\text{o}}$ is attributed to
these contacts (row~8).
The model is consistent with the simulations in all of these trends.
\par
$^{(\text{B.6})}$%
%The fraction of sliding contacts $\eta=15\%$ was imposed as the fifth
%constraint in the entropy model (Eq.~\ref{eq:Gamma7}).
%Although somewhat larger than the 11.2\% for DEM simulations with a
%friction coefficient $\mu$ of 0.50, the model's value is an average
%of the range found in DEM simulations with different coefficients $\mu$.
%, forcing
%agreement with the DEM results in this regard (row~11).
The DEM simulations show that the \emph{directions}
of the contact movements
do not always conform to those expected of an affine deformation field.
As shown in Fig.~\ref{fig:Conventions},
two particles oriented in the
first quadrant would move in a conforming, 
``forward'' direction
if particle ``$j$'' moves upward and to the left over
particle ``$i$''.
Although most contact slip does occur in this forward direction,
the simulations show that
about 40\% of sliding contacts
slip in the \emph{reverse} direction (rows~12--13).
The model predicts this same trend
of a large fraction of contacts slipping
in the reverse direction.
\par
\begin{figure}
\centering
\mbox{%
%  \subfigure[]{\includegraphics%
%  {../Figures/Plot_6_paper_Circles-bidisperse-676-a_0500.pdf}}
%  \ \quad
%  \subfigure[]{\includegraphics%
%  {../Figures/Plot_14_paper_Circles-bidisperse-676-a_0500.pdf}}}
   \includegraphics{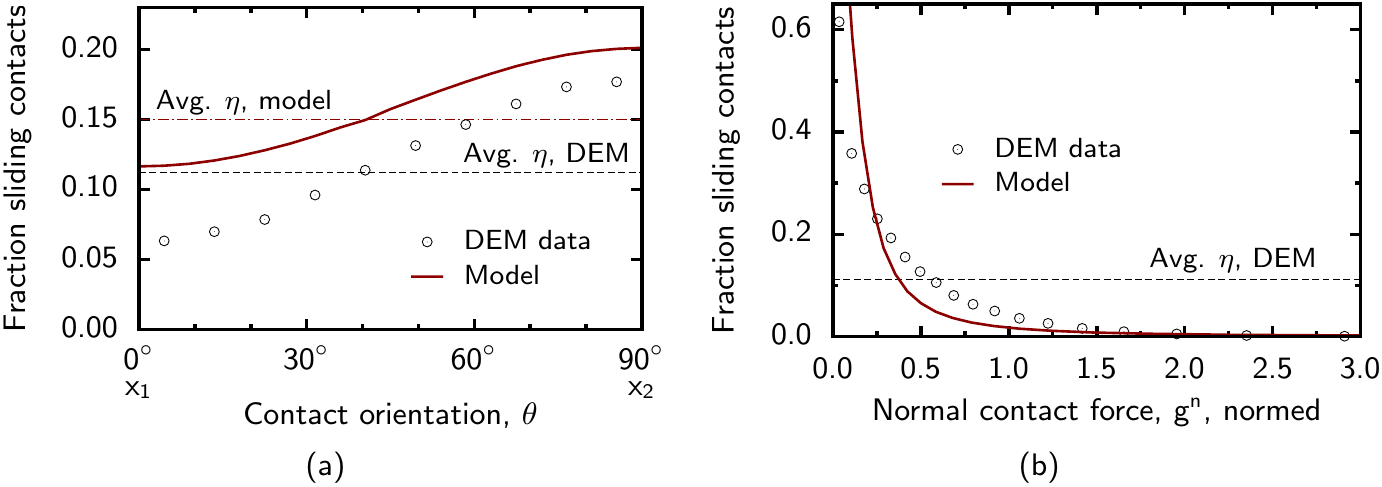}}
\caption{Fraction of sliding contacts as a function of
         (a) contact orientation,
         $\langle 1|g^{\text{t}}\in\{-\mu g^{\text{n}},\mu g^{\text{n}}\}\rangle(\theta^{\text{c}})$, and
         (b) normal contact force,
         $\langle 1|g^{\text{t}}\in\{-\mu g^{\text{n}},
                                      \mu g^{\text{n}}\}\rangle(g^{\text{n}})$.
         \label{fig:results6and14}}
\end{figure}

$^{(\text{B.7})}$%
The likelihood that a contact is sliding
depends upon its orientation and also upon the magnitude of its normal force.
Figure~\ref{fig:results6and14}a
shows that contact slip is most likely among contacts
with normals $\mathbf{n}_{k}$ that are oriented in the
direction of extension (direction $x_{2}$ for the biaxial loading
conditions of this study, Fig.~\ref{fig:Conventions}).
Although the model's results in~\ref{fig:results6and14}a
are shifted because of a
different $\eta$ value, the same trend is predicted. % by the
% model.
%
\par
$^{(\text{B.8})}$%
Figure~\ref{fig:results6and14}b shows the relationship
between the normal contact force
$g^{\text{n}}$ and the fraction of sliding contacts.
With both model and simulations, 
the prevalence of contact sliding decreases
with an increasing contact force: contact slip is far more likely
for lightly loaded contacts than for those bearing a large normal
force.
This trend is also captured by measuring the average normal
forces among the sliding and non-sliding contacts
(Table~\ref{table:Results1}, rows~15--17): the average normal
force among sliding contacts is less than a third of that among
non-sliding contacts.
Moreover, the normal force among forward sliding contacts is 
slightly greater than
the force among contacts sliding in the reverse direction.
All of these trends are predicted with the model.
\par
\begin{figure}
\centering
\mbox{%
%  \subfigure[]{\includegraphics%
%  {../Figures/Plot_16b_paper_Circles-bidisperse-676-a_0500.pdf}}
%  \ \quad
%  \subfigure[]{\includegraphics%
%  {../Figures/Plot_16a_paper_Circles-bidisperse-676-a_0500.pdf}}}
   \includegraphics{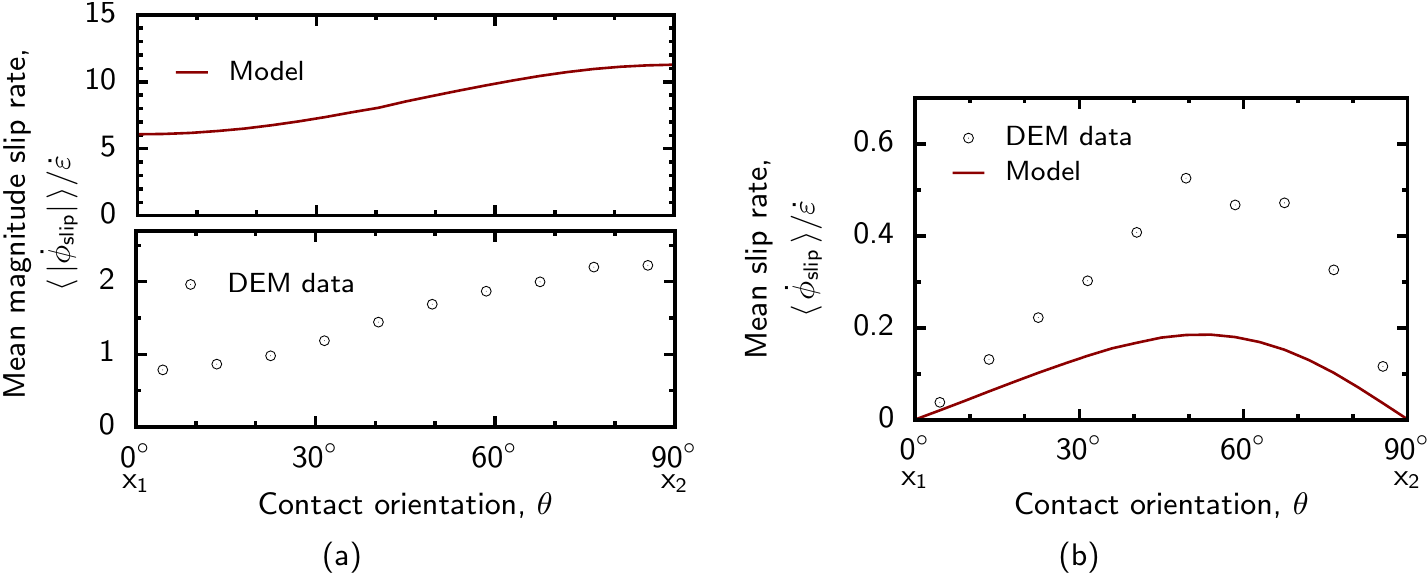}}
\caption{Sliding rate $\dot{\phi}_{\text{slip}}$ and contact
         orientation:
         (a) average magnitude of sliding rate,
             $\langle |\dot{\phi}_{\text{slip}}|/\dot{\varepsilon}
              \rangle(\theta^{\text{c}})$, and
         (b) mean sliding rate,
             $\langle \dot{\phi}_{\text{slip}}/\dot{\varepsilon}
              \rangle(\theta^{\text{c}})$.
         \label{fig:results16}}
\end{figure}

$^{(\text{B.7,B.9})}$%
The sliding (slip) rate $\dot{\phi}_{\text{slip}}$ also depends upon
contact orientation
(Fig.~\ref{fig:results16}).
As with the DEM data, the model predicts a more vigorous
\emph{magnitude} of the slip rate among
contacts that are oriented in the direction of extension
(Fig.~\ref{fig:results16}a).
Although the model is consistent with the DEM data
in this trend,
the magnitudes of the slip rate in the model are considerably
greater than those of the simulations:
the model greatly over-predicts the vigor of contact activity.
Figure~\ref{fig:results16}b shows the \emph{mean}
slip rate as a function of contact orientation.
Consistent with the DEM data,
the model predicts that the mean
slip rate is, on average, in the ``forward''
(positive) direction and that the mean is largest at orientations
oblique to the directions of compression and extension (at
an angle of about 50$^{\circ}$).
\par
\begin{figure}
  \centering
  \includegraphics{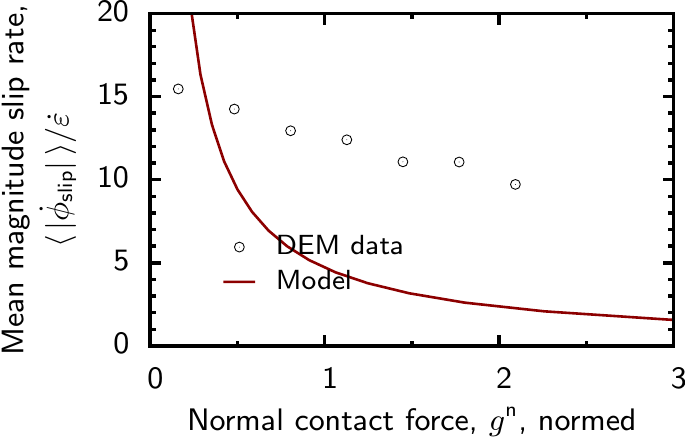}
  \caption{Sliding rate $\dot{\phi}_{\text{slip}}$ and contact
         normal force,
         $\langle |\dot{\phi}_{\text{slip}}|/\dot{\varepsilon}
          \rangle(g^{\text{n}})$.
         \label{fig:results13}}
\end{figure}

$^{(\text{B.8})}$%
Figure~\ref{fig:results13} shows the 
relationship between the contact sliding
rate and the contact normal force.
Although quantitative agreement is poor,
the simulations and the model show a reduction in the
magnitude of slip movements with increasing normal force.
\par
$^{(\text{B.10})}$%
Past DEM analyses of strains within particle clusters have
shown that dilation is associated with clusters that are
elongated in the direction of compression; whereas, clusters
that are elongated in the direction of extension tend to
contract \citep{Nguyen:2009a}.
With the model,
this aspect of granular deformation 
was studied by extracting the dilation rate predicted
% with density $p(\cdots )$
when integration is restricted to
different contact orientations $\theta^{\text{c}}$.
The deformation function $\mathbf{\Gamma}_{\mathbf{L}}$ 
of Eq.~(\ref{eq:GammaL}) was analyzed with the methods
of \ref{app:Conventions} (see Eq.~\ref{eq:avggn}) to compute
the average dilation rate attributed to contacts of any given
orientation $\theta^{\text{c}}$: the average
$\langle \text{tr}(\mathbf{\Gamma}_{\mathbf{L}})/\dot{\varepsilon}
 \rangle(\theta^{\text{c}})$.
\begin{figure}
  \centering
  \includegraphics{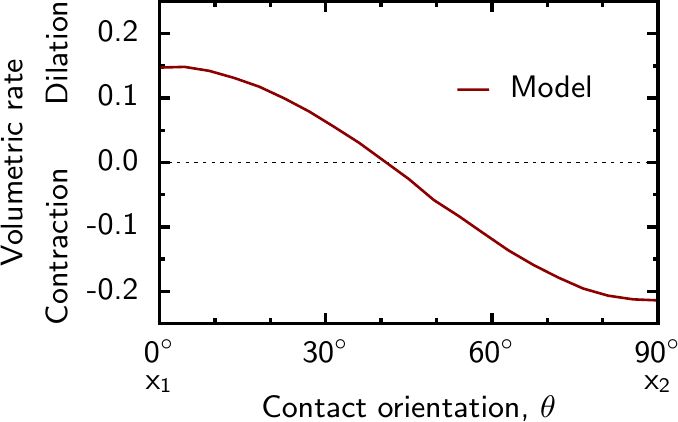}
  \caption{Average volumetric deformation rate attributed to contacts with
         orientation $\theta^{\text{c}}$: the expectation
         $\langle \text{tr}(
            \mathbf{\Gamma}_{\mathbf{L}})/\dot{\varepsilon}
            \rangle(\theta^{\text{c}})$.
         \label{fig:results24}}
\end{figure}
Although the relationship between contact orientation
and dilation was not computed for the current DEM simulations,
the model results, shown in Fig.~\ref{fig:results24}, do confirm past
observations: contacts oriented in the direction of compression
(direction $x_{1}$) are associated with the 
dilation of an assembly;
whereas those oriented in the direction of extension tend
to produce contraction.
\par
\begin{figure}
  \centering
  \includegraphics{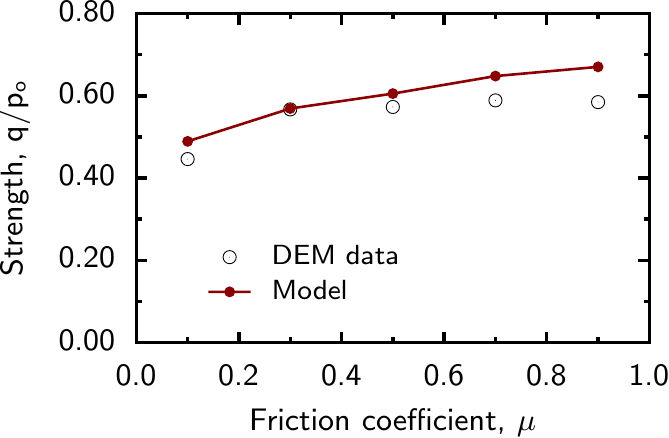}
  \caption{Effect of inter-particle friction coefficient
           $\mu$ on the deviatoric stress during
           critical state flow.
           \label{fig:resultsqmu}}
\end{figure}

$^{(\text{B.12})}$%
As the final aspect of behavior, we consider the effect of 
the inter-particle friction coefficient $\mu$ on bulk strength
at the critical state.
DEM simulations were conducted with five coefficients and the model
was solved with these same values.
Figure~\ref{fig:resultsqmu} compares the model's predictions of
strength with the results of DEM simulations,
expressed as the deviator stress ratio $q/p_{\text{o}}$.
Although the model over-predicts strength at the critical state,
its results follow a similar trend of the DEM data:
strength increases with an increasing friction coefficient,
but the increase is rather small for coefficients greater than 0.30.
% The model and simulations both show that strength is fairly
% insensitive to the coefficient $\mu$.
% The general trend of the model, however, is a \emph{decrease}
% in strength with increasing $\mu$, which differs from the simulations.
% This prediction reflects two counteracting influences.
% When only $\mu$ is increased, the model predicts a larger strength.
% Reducing the sliding fraction $\eta$, however,
% greatly reduces the predicted strength and offsets
% the influence of $\mu$.
% We also see that the close agreement of the
% deviatoric stress ratio $q/p_{\text{o}}$
% when $\mu=0.50$ is somewhat coincidental 
% (row~2 of Table~\ref{table:Results1}).
%
\section{Discussion}\label{sec:Conclusion}
In the Introduction, we listed several micro-scale 
characteristics of critical state granular flow
that have been observed in
past experiments and simulations.
These observations as well as new DEM simulations
were used to evaluate a proposed model for granular flow.
Nearly all of the model's predictions are in qualitative 
(if not quantitative) agreement with observed characteristics.
The model favorably predicts anisotropies of the contact orientations,
contact forces, and contact movements and
of the orientations of those contacts
undergoing slip.
The model also favorably predicts 
relationships between the contact force magnitude, contact motion,
and contact slip direction.
Although several pages were required for its derivation,
and its solution requires quite demanding computations,
the model's concept is fairly simple:
the contact landscape of motion, force, and orientation
is maximally disordered at the critical state,
after one accounts for the
biases that arise from certain fundamental and empirical
aspects of flow.
\par
The model adopts individual contacts
as its generic units, but without identifying
their locations or their affiliations with other contacts.
As such, it is certainly unable to resolve the spatial
localization and meso-scale
patterning that is usually found within granular flows
(item D.1 in the Introduction).
%This inability also discounts 
Past studies have shown that 
motions and forces are spatially coordinated (correlated) over distances
of several particle diameters
%\citep{Mueth:2000a,Kuhn:2003d}.
\citep{Kuhn:2003d}.
Such patterning and correlation 
affirm a greater order (and lower entropy)
than is accessible with the model.
% Compared with an entirely isotropic state with uniform distributions
% of force and motion --- a more ordered, certain, and uniform condition ---
% the model captures the more disordered condition of flow.
%of lower disorder (lower entropy) and 
% and the diminished uniformity (i.e., higher entropy) of granular flow.
\par
Biased by the loading direction in the second constraint,
the model predicts a maximally disordered condition that is anisotropic.
Although anisotropy, by itself, connotes greater order than isotropy,
the small increase in the order of contact orientation is more
than offset by an increase in the disorder associated with the
contact forces and movements. % that are in keeping with
% the imposed constraints.
\par
In some respects,
the model either poorly predicts
behavior or requires an additional constraint
to force agreement.
These shortcomings include the density of the tangential contact
forces among non-sliding contacts (Fig.~\ref{fig:results9and11}b) and
the fraction of sliding contacts (constraint~5).
% and the vigor of contact movements (Fig.~\ref{fig:results16}a).
In other respects, the model is consistent with trends
in the simulation results but is
% Many other results 
in poor \emph{quantitative} agreement % with the DEM simulations.
(for example, in the vigor of the contact movements,
Fig.~\ref{fig:results16}a).
Although its successes are promising,
the model's shortcoming are more revealing, raising this question:
why are some behaviors not amenable to a simple,
contact-centric statistical
treatment?
\par
An underlying assumption of the Jaynes maximum entropy
(MaxEnt) approach is that each micro-state is equally
probable, provided that the micro-state is consistent with
available information (i.e., the imposed constraints).
In the current setting, each of the ``$k$'' micro-states
$\{\cdots\}$ in the dimension-$M^{6}$
phase space belongs to a particular macro-state $(\cdots )$
and is assumed equiprobable with the other micro-states
that populate the same macro-state.
In the model,
the micro-states only incorporate contact information
and are devoid of content on
the particles' locations or their topologic arrangement.
The equiprobable assumption ignores the fundamental meso-scale
restrictions that are shared by the particles and contacts:
the contact forces on each particle must be in equilibrium,
and the contact movements must be compatible
% with each other and 
with
a corresponding set of \emph{particle} movements and rotations.
These additional restrictions operate at a meso-scale
much larger than individual contacts,
and will bias some micro-states
and disallow others altogether, 
creating an uneven landscape of
the dimension-$M^{6}$ phase space of micro-states.
For example, without the rather coersive
restriction on the fraction of sliding contacts (constraint~5),
the model predicts a fraction greater than 80\%, rather
than the 11.2\% measured in the simulations.
This result suggests that particle motions are coordinated
(ordered) in a manner that greatly reduces the population
of sliding contacts.
Reasoning of a fundamental nature,
applied to meso-scale conditions,
is palpably needed to replace
brute force,
empirical constraints, such as our constraint on the
fraction of sliding contacts
and other constraints that are approximate in nature
(constraints~3 and~4).
We consider the lack of meso-scale information the
primary deficiency of the model,
and a means of introducing such information is
suggested in the following paragraphs.
On the other hand,
those micro-scale characteristics and trends that are favorably
predicted are apparently insensitive to meso-scale
equilibrium and compatibility limitations,
but are instead determined (or biased) by the rather broad
constraints of pressure, of dissipation consistency,
and of the isochoric condition.
% Such favorably predicted characteristics remind one
% of Boltzmann's H-theorem of gasses, in which
% the distribution of the molecules' velocities
% is insensitive to details of their collisions.
% Better predictions can
% be achieved, of course, %(and have been achieved, in some limiting cases) 
% by fully respecting the equilibrium or
% compatibility restrictions.
% As an example,
% more realistic results have been obtained for the probability density
% of the normal contact forces,
% although for non-flowing or for frictionless materials
% \citep{Coppersmith:1996a,Snoeijer:2004a,Metzger:2008a,Chakraborty:2010a}.
\par
% In constructing the current model, its
% Four of the model's five constraints were based upon
% fundamental principles, but a simple four-constraint 
% model over-predicts the fraction of sliding
% particles and the vigor of contact movements.
% A four-constraint model, however, does predict a consistent
% increase in strength when the friction coefficient $\mu$ is increased,
% a favorable trend that is negated when the additional
% auxiliary constraints are included in the model (Fig.~\ref{fig:resultsqmu}).
% Although the model's shortcomings are likely due to its
% local, contact-centric approach,
% The author is unaware of
% a fundamental (non-empirical) rationale that would
% place proper limits on the contact movements, other than
% the rather \emph{ad hoc}, direct approach of applying empirical information,
% as with the fifth, auxiliary constraint.
% (or a direct constraint on contact sliding, as with the modified model).
% Reasoning of a more fundamental nature 
% is palpably needed to replace the
% model's rather inelegant, brute force application of 
% empirical constraints.
% In this regard, insight might be gained 
% from a meso-scale analysis of the equilibrium
% and compatibility within particle clusters.
%But as difficulty as such analysis
%seems, it is likely more difficult than it seems.
%
% An improved entropy model of steady frictional
% flow 
% Improvements can be obtained, of course, by incorporating additional
% information.
\par
The paper makes exclusive use of information that is formulated
as moment constraints on the density $p(\cdots)$ (Eqs.~\ref{eq:GammaInt} 
and~\ref{eq:Constraints}).
An alternative form of information was introduced by the
author in a paper on topologic entropy \citep{Kuhn:2014a}.
In this work, a relative entropy principle
(cross-entropy or Kullback-Leibler entropy) was applied,
admitting information in the form of
certain \emph{a priori} inclinations
(priors) of $p(\cdots )$.
Because these inclinations can have either an empirical or theoretical
origin,
this approach allows information --- even imperfect information ---
gained from one theory
(for example, an affine field approach)
% theoretical distributions of contact force from a 
% frictionless non-flow model
% that respects equilibrium)
to be included in a
shear-driven entropy model.
\par
We should also consider the entropy model in
relation to mean-field upscaling models,
which have been used with reasonable success in estimating the 
small-strain stiffness of dense granular materials.
These models can be more descriptive than
predictive, as they require extensive additional information,
in the form of contact stiffnesses, packing data, etc.
Mean-field models, however, do provide a clearer,
deterministic connection
between the micro-scale and bulk behaviors:
the micro and macro behaviors are unambiguously linked through
contact rules. % and an affine, mean-field assumption.
The corresponding connection is rather opaque 
in the paper's approach.
For example,
%how do the model's constraints
%lead to a certain fraction of sliding contacts, $\eta$?
why do the constraints lead to a
greater number of forward sliding contacts than
reverse sliding contacts?
What features of the model lead to smaller normal forces
among sliding contacts than for non-sliding contacts?
Although the model yields some convincing predictions,
the model is certainly 
enigmatic in regard to its results,
as the model's predictions are extracted only through
a painstaking integration of distribution averages.
\par
Finally, we suggest future extensions of this work.
With some difficulty,
the model could certainly be extended to three dimensions,
which would require that the
tangential forces and motions be expressed with Euler angles rather
than the single $\theta$ orientation of the paper's two-dimensional
setting.
The model could also be modified for non-circular or non-spherical
particles, by allowing different orientations of the contacts and the
branch vectors, $\theta^{\text{c}}$ and $\theta^{\ell}$ %, to differ
(i.e., by relaxing the Dirac restriction).
In the Introduction, we had noted that
granular materials consistently converge toward the critical state when
loaded from differing initial conditions.
Although the current model addresses the terminal condition
% and certain measures of disorder
at the critical state,
the disorder in the paper's contact quantitites (and other quantities, as well)
% the current and possibly other measures of disorder
%A measure of entropy, perhaps that of Eq.~(\ref{eq:H}), 
could be tracked during the evolving condition of granular loading.
An ideal entropy measure would increase in a consistent, monotonic
manner during the loading process.
This approach might lead to an improved entropy definition %that
based upon a more appropriate phase space.
% would include a broader range of micro-scale activity and disorder: 
% changes in the
% topology and geometric arrangement of particles as well as
% the contact
% characteristics that are the exclusive focus of the current work.
%
%\par
%Kruyt acknowledgement.
%\par
%Satake memorial.
%
%
\section*{\normalsize Acknowledgement}
The author greatfully acknowledges productive discussions
of this work with Dr. Niels Kruyt.
\section*{\normalsize Dedication}
The paper is dedicated to the memory of Dr. Masao Satake
(1927--2013), who made significant fundamental contributions
to granular mechanics.
\appendix
\section{Integrations}\label{sec:Integrations}
Each constraint involves integrating a product 
$\Gamma_{i}(\cdots)p(\cdots)$, as
sketched in Eq.~(\ref{eq:GammaInt}).
Any evaluation of these integrals must, in practice, 
reconcile the non-smooth
nature of these integrands:
both the rigid-frictional constitutive constraint of 
Eq.~(\ref{eq:GtRestrict}) and the kernel
$K_{1}()$
are discontinuous.
In the dimension-6 phase space,
large non-zero values of 
$\dot{\phi}_{\text{slip}}$ can exist alongside
an enforced zero value at the juncture 
of sliding and non-sliding
behaviors: at hyper-planes $g^{\text{t}}=\pm \mu g^{\text{n}}$.
This difficulty is resolved by splitting the original integral
into three parts, so that each part addresses a single branch
of the three constitutive cases in Eq.~(\ref{eq:GtRestrict}):
\begin{align}
\underset{( g^{\text{n}},g^{\text{t}}, 
        \theta^{\text{c}},\theta^{\ell},
        \dot{\phi}_{\text{slip}},\dot{\phi}_{\text{rigid}} )}
        {\idotsint}=    \label{eq:SplitIntegrals}
%\\ \notag
& \quad%\quad\quad\quad
\underset{\substack{\mathbb{R}^{+}\\g^{\text{n}} }}{\int}\;
\underset{\substack{[0,\,2\pi)\\ \theta^{\text{c}} }}{\int}\;
\underset{\substack{[0,\,2\pi)\\ \theta^{\ell} }}{\int}\;
\underset{\substack{\mathbb{R}^{-}\\ \dot{\phi}_{\text{slip}} }}{\int}\;
\underset{\substack{\mathbb{R}\\ \dot{\phi}_{\text{rigid}} }}{\int}\;
\begin{array}{l}
  \text{ with }g^{\text{t}}=-\mu g^{\text{n}}\\
  \text{ and }\dot{\phi}_{\text{slip}} < 0
\end{array}
\\ \notag
& \quad%\quad\quad
+
\underset{\substack{\mathbb{R}^{+}\\g^{\text{n}} }}{\int}\;
\underset{\substack{(-1,1)\\g^{\text{t}}}}
  {\int}\;
\underset{\substack{[0,\,2\pi)\\ \theta^{\text{c}} }}{\int}\;
\underset{\substack{[0,\,2\pi)\\ \theta^{\ell} }}{\int}\;
\underset{\substack{\mathbb{R}\\ \dot{\phi}_{\text{rigid}} }}{\int}
\begin{array}{l}
\text{ with }g^{\text{t}}\in(-\mu g^{\text{n}},\mu g^{\text{n}})\\
\text{ and }\dot{\phi}_{\text{slip}} = 0
\end{array}
\\ \notag
& \quad%\quad\quad
+
\underset{\substack{\mathbb{R}^{+}\\g^{\text{n}} }}{\int}\;
\underset{\substack{[0,\,2\pi)\\ \theta^{\text{c}} }}{\int}\;
\underset{\substack{[0,\,2\pi)\\ \theta^{\ell} }}{\int}\;
\underset{\substack{\mathbb{R}^{+}\\ \dot{\phi}_{\text{slip}} }}{\int}\;
\underset{\substack{\mathbb{R}\\ \dot{\phi}_{\text{rigid}} }}{\int}\;
\begin{array}{l}
  \text{ with }g^{\text{t}}=\mu g^{\text{n}}\\
  \text{ and }\dot{\phi}_{\text{slip}} > 0
\end{array}
\end{align}
The second integral on the right addresses the non-sliding branch, and
the first and third integrals address sliding in the ``reverse''
and ``forward'' directions (sliding directions
that are contrary to and consistent with those of
affine deformation, as described in Section~\ref{sec:variables}).
Note that the three integrals are 5-fold, compared
with the original 6-fold integrals (on the left
of Eq.~\ref{eq:SplitIntegrals}),
greatly reducing the complexity of
numerical evaluations.
Separating (\ref{eq:SplitIntegrals}) as three integrals
also permits a direct application of 
constraint~5 (Eq.~\ref{eq:Gamma7}) 
and aids in computing separate statistics for the
non-sliding and sliding (both forward and reverse) contacts
%These statistics are among those presented in 
(Section~\ref{sec:compare1}).
%
%and this observation will
%aid in evaluating certain combined integrands that accompany
%the maximum entropy condition.
%
%
\section{Entropy numerics}\label{sec:Numerics}
The entropy $H$ in Eqs.~(\ref{eq:H})--(\ref{eq:MaxZ})
is maximized by finding the five multipliers
$\lambda_{i}$ that satisfy the constraints
$\langle\Gamma_{i}(\cdots)\rangle=\overline{\Gamma}_{i}$
of Eq.~\ref{eq:Constraints}.
%of Eqs.~(\ref{eq:Gamma1}), (\ref{eq:Gamma2}), (\ref{eq:Gamma3}),
%and~(\ref{eq:Gamma4}).
That is, the proper $\lambda_{i}$ are the roots of five equations
\begin{equation}
f_{i}(\boldsymbol{\lambda}) =
A_{i}(\boldsymbol{\lambda})
- \overline{\Gamma}_{i} Z(\boldsymbol{\lambda}) = 0
\end{equation}
where argument $\boldsymbol{\lambda}$ represents the list
$\{\lambda_{1}, \lambda_{2},\ldots,\lambda_{5}\}$, and
\begin{equation}
A_{i}(\boldsymbol{\lambda}) =
\underset{( g^{\text{n}},g^{\text{t}}, 
        \theta^{\text{c}},\theta^{\ell},
        \dot{\phi}_{\text{slip}},\dot{\phi}_{\text{rigid}} )}
        {\idotsint}
  \Gamma_{i}(\cdots)
  \exp\left(
  -\sum_{j=1}^{5}
\lambda_{j}\Gamma_{j}(\cdots)
\right)
\end{equation}
%
%For the simplified model with its three constraints, argument
%$\boldsymbol{\lambda}=\{\lambda_{1},\lambda_{2},\ldots,\lambda_{7}\}$.
We used the \texttt{minpack} library to solve these equations
by minimizing the sum of their squared residuals
(specifically, function \texttt{hybrj1}, which applies
the Powell hybrid method) \citep{More:1984a}.
The solver requires evaluation of the partition function
$Z(\boldsymbol{\lambda})$, 
the five functions $f_{i}(\boldsymbol{\lambda})$,
and the $5\times 5$ Jacobian
\begin{align}
\frac{\partial f_{i}}{\partial\lambda_{j}} &= 
  -B_{ij} + \overline{\Gamma}_{i}A_{j}\\
B_{ij} &=
\underset{( g^{\text{n}},g^{\text{t}},
        \theta^{\text{c}},\theta^{\ell},
        \dot{\phi}_{\text{slip}},\dot{\phi}_{\text{rigid}} )}
        {\idotsint}
  \Gamma_{i}(\cdots)\Gamma_{j}(\cdots)
  \exp\left(
  -\sum_{k=1}^{5}
\lambda_{k}\Gamma_{k}(\cdots)
\right)
\end{align}
The total of $1+5+25=31$ integrands were evaluated within
each iteration of the solver \texttt{hybrj1}.
Each \emph{symbolic} integral is itself
the sum of three integrals, each 5-fold
(see Eq.~\ref{eq:SplitIntegrals}).
%One integral of each pair is of dimension 5 and spans
%the full domain 
%$\{\cdot,\cdot,\theta^{\text{c}},\theta^{\ell},\cdot,\cdot\}$;
%the other integral of the pair is of dimension 4 and spans
%the reduced domain
%$\{\cdot,\cdot,\theta^{\text{c}},\theta^{\text{c}},\cdot,\cdot \}$,
%as in Eq.~(\ref{eq:IntegralsDirac}) and its discussion.
%Evaluating these six integrals is particularly vexing because
%of their discontinuous character, as described in the beginning
%of Section~\ref{sec:Integrations}.
%With the three constraints of the simplified model,
%each integral must be evaluated with $1+3+9=10$ integrands.
All integrals were evaluated with the \texttt{CUBA} library
(specifically the function \texttt{llcuhre}, which applies
adaptive polynomial cubature rules and
permits a vector of integrands to be evaluated
with each call). % \citep{Haha:2005a}.
Except for the variables
$g^{\text{t}}$, $\theta^{\text{c}}$, and $\theta^{\ell}$,
which have ranges of 
$[-\mu g^{\text{n}},\mu g^{\text{n}}]$ or $[0,2\pi )$,
all integrations are improper
with infinite range.
A change of variable, for example $u=\tan^{-1}g^{\text{n}}$, 
was applied in such cases, and integrations were evaluated with the
following support:
%ranges were set as follows:
$g^{\text{n}}\in [0,10]$,
$\dot{\phi}_{\text{slip}}\in [-200,200]$,
and $\dot{\phi}_{\text{rigid}}\in [-200,200]$.
Two hundred million points were queried in evaluating each
integrand within each of the
three integrals within each iteration.
The mean stress $p_{\text{o}}$ and strain rate $\dot{\varepsilon}$
were both set to 1.0, and the factor $M^{\text{int}}\overline{\ell}^{2}/A$
was 1.50. %, as derived in Appendix~\ref{sec:Counting}.
\par
After solving the multipliers $\lambda_{i}$, meaningful results,
such as those in Section~\ref{sec:compare1}
must be extracted by evaluating integrals with appropriate
integrands to compute the relevant $\langle \cdot \rangle$ quantities.
These calculations are described in the next appendix.
\section{Statistical calculations and notation}\label{app:Conventions}
After determining the multipliers $\lambda_{i}$,
meaningful information can be extracted from the probability
density $p(\cdots )$ by post-processing
of various integrals in the form of Eq.~(\ref{eq:SplitIntegrals}).
Statistical notations used in the paper are intended
to suggest their numerical evaluation, although some 
differ from traditional notations.
The expected value of a function $h(\cdots)$ is written as
$\langle h(\cdots)\rangle$ or $\langle h\rangle$
and is computed, as in Eqs.~(\ref{eq:GammaInt})
and~(\ref{eq:SplitIntegrals}),
by integrating across the full domain
$( g^{\text{n}},g^{\text{t}},\theta^{\text{c}},\theta^{\ell},
 \dot{\phi}_{\text{slip}},\dot{\phi}_{\text{rigid}} )$.
When the domain is
restricted by a condition $\Psi$, we use the notation
$\langle h(\cdots)\rangle |_{\Psi}$,
\begin{equation}
\langle h(\cdots)\rangle\,|_{\Psi} = 
\left.
\underset{( g^{\text{n}},g^{\text{t}}, 
        \theta^{\text{c}},\theta^{\ell},
        \dot{\phi}_{\text{slip}},\dot{\phi}_{\text{rigid}} )}
        {\idotsint} 
\!\!\!\!\! h(\cdots)\,  p(\cdots)\right|_{\Psi}
\label{eq:CondProp}
\end{equation}
in which the integrand is zero outside of this restricted
domain (i.e., where condition $\Psi$ is not met).
For example, the probability of condition $\Psi$ is simply
$\langle 1\rangle |_{\Psi}$, with $h=1$.
%Likewise, integration of the Dirac integral
%$K_{2}( \theta^{\text{c}},\theta^{\ell})$ 
%is equivalent to the restriction 
%$\Psi \Rightarrow \theta^{\text{c}}=\theta^{\ell}$,
%
%\begin{equation}
%\langle K_{2}( \theta^{\text{c}},\theta^{\ell})\,f(\cdots)\rangle
%= \langle f(\cdots)\rangle |_{\theta^{\text{c}}=\theta^{\ell}}
%\end{equation}
%
%as in Eq.~(\ref{eq:IntegralsDirac}).
The conditional expectation of function $h(\cdots)$ subject
to $\Psi$ is written as $\langle h(\cdots)|\Psi\rangle$ and is
computed as
\begin{equation}
\langle h(\cdots)|\Psi\rangle=
\frac
  {\langle h(\cdots)\rangle |_{\Psi}}
  {\langle 1\rangle |_{\Psi}}
\end{equation}
We also compare the statistics of contacts that are not sliding
or are sliding in the ``reverse'' or ``forward'' directions.
The three cases correspond to different $\Psi$ conditions:
$\dot{\phi}_{\text{slip}}=0$, $\dot{\phi}_{\text{slip}}=<0$, and
$\dot{\phi}_{\text{slip}}=>0$
(or alternatively,
$g^{\text{t}}\in(-\mu g^{\text{n}},\mu g^{\text{n}})$,
$g^{\text{t}}=-\mu g^{\text{n}}$, and $g^{\text{t}}=\mu g^{\text{n}}$),
as explicitly isolated
with the three integrals in Eq.~(\ref{eq:SplitIntegrals}).
\par
The marginal probability density of
a single contact quantity will be written as a derivative,
in keeping with the manner in which it is computed.
For example, the marginal probability density of 
the contact orientation $\theta^{\text{c}}$
is written as the derivative
$\mathcal{D}_{\theta^{\text{c}}} \langle 1\rangle$:
\begin{equation}
\mathcal{D}_{\theta^{\text{c}}}\langle 1 \rangle
=\underset{\Delta\theta\rightarrow 0}{\lim}
\frac
{\langle 1 \rangle
   |_{\theta^{\text{c}}<\theta<\theta^{\text{c}} + \Delta\theta}}
{\Delta\theta}
\end{equation}
As another example, the average expected normal force $g^{\text{n}}$ 
%of function $f(\cdots)=g^{\text{n}}$ 
as a function of contact orientation
$\theta^{\text{c}}$ is written and computed as
\begin{equation}
\langle g^{\text{n}} \rangle (\theta^{\text{c}}) =
\frac{\mathcal{D}_{\theta^{\text{c}}}\langle g^{\text{n}}\rangle}
     {\mathcal{D}_{\theta^{\text{c}}}\langle 1\rangle} =
\underset{\Delta\theta\rightarrow 0}{\lim}
\frac
{\langle g^{\text{n}} \rangle
   |_{\theta^{\text{c}}<\theta<\theta^{\text{c}} + \Delta\theta}}
{\langle 1 \rangle
   |_{\theta^{\text{c}}<\theta<\theta^{\text{c}} + \Delta\theta}}
\label{eq:avggn}
\end{equation}
\section{DEM simulations, continued}\label{app:DEM}
Multiple discrete element (DEM) simulations of biaxial
compression were conducted on
bi-disperse assemblies of 676 particles.
The two disk varieties have ratios
of 1.5:1 in size, 1:2.25 in number, and 1:1 in cumulative area
(that is, 468 particles of size $1.0$ and 208 of size $1.5$).
The assemblies were small enough 
to prevent gross non-homogeneity
in the form shear bands
(no bands were observed, by using the
methods of \citealp{Kuhn:1999a}),
yet large enough to capture the average, bulk material behavior.
To develop more robust statistics, 
168 different assemblies were created
by compacting random sparse frictionless mixtures of the two disk sizes
into dense isotropic packings within periodic boundaries until
the average contact indentation was 
$0.0002$ times the average radius.
In the subsequent biaxial loading sequences,
linear contact stiffnesses were applied between particles
with equal tangential and normal coefficients
($k^{\text{t}}=k^{\text{n}}$), and
the base friction coefficient $\mu=0.5$ 
was enforced during the pair-wise
particle interactions~\citep{daCruz:2005a}.
Coefficients $\mu=0.1$, 0.3, 0.7, and 0.9 were used in
separate simulations, all starting from the same 168 assemblies.
Using the standard 
DEM algorithm,
the initially square assemblies were horizontally
compressed in increments $\Delta\epsilon_{11}=1\times10^{-6}$
while maintaining a constant mean stress of 
$2\times 10^{-4}k^{\text{n}}$ \citep{Cundall:1979a}.
The strain rate was sufficiently slow to maintain
the quasi-static condition, with the inertial number
$I$ equal to $2.5\times 10^{-4}$
(i.e., the ratio of the shear time $1/\dot{\varepsilon}$
to the inertial time of a particle $\sqrt{m/p_{\text{o}}}$,
see \citealp{daCruz:2005a}).
These loading conditions coincide with those described in
Sections~\ref{sec:isochoric}--\ref{sec:auxiliary}
and result in the
stress, fabric, and volumetric behavior that are shown in
Fig.~\ref{fig:crs}.
A large initial stiffness causes the deviatoric
stress to rise quickly from zero to a peak stress at
strain $-\varepsilon_{11}=2\%$,
and the critical state condition is
attained at compressive strains $-\varepsilon_{11}$ of 16--18\%.
During subsequent steady-state deformation, 
the contact conditions were 
interrogated at five strains between 
$-\varepsilon_{11}=16\%$ and~$25\%$ (Fig.~\ref{fig:crs}),
and rates $\dot{n}$ and $\dot{\phi}$ were determined with pairs
of interrogations separated by strain 
$\Delta\varepsilon_{11}=1\times 10^{-5}$.
Applying the ergodicity principle, micro-state statistics 
were averaged
across the five strains and the 168 assemblies, 
involving 840 pairs of interrogations
containing about 820,000 contacts.
Another series of simulations were run with poly-disperse
assemblies having particle sizes spanning a range of 3.0.
The results were similar to those of the bi-disperse assemblies,
and only the latter are reported herein.
\section*{References}
\bibliographystyle{elsarticle-harv}
%\bibliography{Kuhn}

\begin{thebibliography}{40}
\expandafter\ifx\csname natexlab\endcsname\relax\def\natexlab#1{#1}\fi
\expandafter\ifx\csname url\endcsname\relax
  \def\url#1{\texttt{#1}}\fi
\expandafter\ifx\csname urlprefix\endcsname\relax\def\urlprefix{URL }\fi

\bibitem[{Bagi(2006)}]{Bagi:2006a}
Bagi, K., 2006. Analysis of microstructural strain tensors for granular
  assemblies. Int. J. Solids Struct. 43~(10), 3166--3184.

\bibitem[{Blumenfeld and Edwards(2009)}]{Blumenfeld:2009a}
Blumenfeld, R., Edwards, S.~F., 2009. On granular stress statistics:
  Compactivity, angoricity, and some open issues. The Journal of Physical
  Chemistry B 113~(12), 3981--3987.

\bibitem[{Brown(2000)}]{Brown:2000a}
Brown, C.~B., 2000. Entropy and granular materials: model. J. Eng. Mech.
  126~(6), 599--604.

\bibitem[{Brown et~al.(2000)Brown, Elms, Hanson, Nikzad, and
  Worden}]{Brown:2000b}
Brown, C.~B., Elms, D.~G., Hanson, M.~T., Nikzad, K., Worden, R.~E., 2000.
  Entropy and granular materials: experiments. J. Eng. Mech. 126~(6), 605--610.

\bibitem[{Chakraborty(2010)}]{Chakraborty:2010a}
Chakraborty, B., 2010. Statistical ensemble approach to stress transmission in
  granular packings. Soft Matt. 6~(13), 2884--2893.

\bibitem[{Cundall and Strack(1979)}]{Cundall:1979a}
Cundall, P.~A., Strack, O. D.~L., 1979. A discrete numerical model for granular
  assemblies. G{\'{e}}otechnique 29~(1), 47--65.

\bibitem[{da~Cruz et~al.(2005)da~Cruz, Emam, Prochnow, Roux, and
  Chevoir}]{daCruz:2005a}
da~Cruz, F., Emam, S., Prochnow, M., Roux, J.-N., Chevoir, F., 2005.
  Rheophysics of dense granular materials: Discrete simulation of plane shear
  flows. Phys. Rev. E 72~(2), 021309.

\bibitem[{Desrues and Viggiani(2004)}]{Desrues:2004a}
Desrues, J., Viggiani, G., 2004. Strain localization in sand: an overview of
  the experimental results obtained in {G}renoble using stereophotogrammetry.
  Int. J. Numer. and Anal. Methods in Geomech. 28, 279--321.

\bibitem[{Goddard(2004)}]{Goddard:2004a}
Goddard, J.~D., 2004. On entropy estimates of contact forces in static granular
  assemblies. Int. J. Solids Struct. 41~(21), 5851--5861.

\bibitem[{H{\"a}rtl and Ooi(2008)}]{Hartl:2008a}
H{\"a}rtl, J., Ooi, J.~Y., 2008. Experiments and simulations of direct shear
  tests: porosity, contact friction and bulk friction. Granul. Matter 10~(4),
  263--271.

\bibitem[{Jaynes(1957)}]{Jaynes:1957a}
Jaynes, E.~T., 1957. Information theory and statistical mechanics. Phys. Rev.
  106~(4), 620--630.

\bibitem[{Jenkins and Strack(1993)}]{Jenkins:1993a}
Jenkins, J.~T., Strack, O. D.~L., 1993. Mean-field inelastic behavior of random
  arrays of identical spheres. Mech. of Mater. 16~(1), 25--33.

\bibitem[{Kruyt and Antony(2007)}]{Kruyt:2007a}
Kruyt, N.~P., Antony, S.~J., 2007. Force, relative-displacement, and work
  networks in granular materials subjected to quasistatic deformation. Phys.
  Rev. E 75~(5), 051308.

\bibitem[{Kruyt and Rothenburg(2014)}]{Kruyt:2014a}
Kruyt, N.~P., Rothenburg, L., 2014. On micromechanical characteristics of the
  critical state of two-dimensional granular materials. Acta Mechanica 225~(8),
  2301--2318.

\bibitem[{Kuhn(1999)}]{Kuhn:1999a}
Kuhn, M.~R., 1999. Structured deformation in granular materials. Mech. of
  Mater. 31~(6), 407--429.

\bibitem[{Kuhn(2003)}]{Kuhn:2003d}
Kuhn, M.~R., 2003. Heterogeneity and patterning in the quasi-static behavior of
  granular materials. Granul. Matter 4~(4), 155--166.

\bibitem[{Kuhn(2004{\natexlab{a}})}]{Kuhn:2004a}
Kuhn, M.~R., 2004{\natexlab{a}}. Boundary integral for gradient averaging in
  two dimensions: application to polygonal regions in granular materials. Int.
  J. Numer. Methods Eng. 59~(4), 559--576.

\bibitem[{Kuhn(2004{\natexlab{b}})}]{Kuhn:2003h}
Kuhn, M.~R., 2004{\natexlab{b}}. Rates of stress in dense unbonded frictional
  materials during slow loading. In: Antony, S.~J., Hoyle, W., Ding, Y. (Eds.),
  Advances in Granular Materials: Fundamentals and Applications. Royal Society
  of Chemistry, London, U.K., pp. 1--28.

\bibitem[{Kuhn(2014)}]{Kuhn:2014a}
Kuhn, M.~R., 2014. Dense granular flow at the critical state: maximum entropy
  and topological disorder. Granul. Matter 16~(4), 499--508.

\bibitem[{Kuhn and Bagi(2004)}]{Kuhn:2004k}
Kuhn, M.~R., Bagi, K., 2004. Contact rolling and deformation in granular media.
  Int. J. Solids Struct. 41~(21), 5793--5820.

\bibitem[{Kuhn and Bagi(2005)}]{Kuhn:2005c}
Kuhn, M.~R., Bagi, K., 2005. On the relative motions of two rigid bodies at a
  compliant contact: application to granular media. Mech. Res. Comm. 32~(4),
  463--480.

\bibitem[{Liao et~al.(1997)Liao, Chang, and Young}]{Liao:1997a}
Liao, C.-L., Chang, T.-P., Young, D.-H., 1997. Stress-strain relationship for
  granular materials based on the hypothesis of best fit. Int. J. Solids
  Struct. 34~(31--32), 4087--4100.

\bibitem[{Majmudar and Bhehringer(2005)}]{Majmudar:2005a}
Majmudar, T.~S., Bhehringer, R.~P., 2005. Contact force measurements and
  stress-induced anisotropy in granular materials. Nature 435~(1079),
  1079--1082.

\bibitem[{Majmudar et~al.(2007)Majmudar, Sperl, Luding, and
  Behringer}]{Majmudar2007a}
Majmudar, T.~S., Sperl, M., Luding, S., Behringer, R.~P., 2007. Jamming
  transition in granular systems. Phys. Rev. Lett. 98~(5), 058001.

\bibitem[{Mor{\'e} et~al.(1984)Mor{\'e}, Sorensen, Hillstrom, and
  Garbow}]{More:1984a}
Mor{\'e}, J.~J., Sorensen, D.~C., Hillstrom, K.~E., Garbow, B.~S., 1984. The
  {MINPACK} project. In: Cowell, W.~R. (Ed.), Sources and Development of
  Mathematical Software. Prentice-Hall, Englewood Cliffs, N.J., pp. 88--111.

\bibitem[{Nguyen et~al.(2009)Nguyen, Magoariec, Cambou, and
  Danescu}]{Nguyen:2009a}
Nguyen, N.-S., Magoariec, H., Cambou, B., Danescu, A., 2009. Analysis of
  structure and strain at the meso-scale in 2d granular materials. Int. J.
  Solids Struct. 46~(17), 3257--3271.

\bibitem[{Pe{\~{n}}a et~al.(2008)Pe{\~{n}}a, Lizcano, Alonso-Marroquin, and
  Herrmann}]{Pena:2008a}
Pe{\~{n}}a, A.~A., Lizcano, A., Alonso-Marroquin, F., Herrmann, H.~J., 2008.
  Biaxial test simulations using a packing of polygonal particles. Int. J.
  Numer. and Anal. Methods in Geomech. 32~(2), 143--160.

\bibitem[{Radjai et~al.(1998)Radjai, Wolf, Jean, and Moreau}]{Radjai:1998a}
Radjai, F., Wolf, D.~E., Jean, M., Moreau, J.-J., 1998. Bimodal character of
  stress transmission in granular packings. Phys. Rev. Lett. 80~(1), 61--64.

\bibitem[{Rothenburg(1980)}]{Rothenburg:1980a}
Rothenburg, L., 1980. Micromechanics of idealized granular systems. Ph.D.
  thesis, Carleton University, Ottawa, Ontario, Canada.

\bibitem[{Rothenburg and Bathurst(1989)}]{Rothenburg:1989a}
Rothenburg, L., Bathurst, R., 1989. Analytical study of induced anisotropy in
  idealized granular materials. G{\'{e}}otechnique 39~(4), 601--614.

\bibitem[{Rothenburg and Kruyt(2009)}]{Rothenburg:2009a}
Rothenburg, L., Kruyt, N.~P., 2009. Micromechanical definition of an entropy
  for quasi-static deformation of granular materials. J. Mech. Phys. Solids
  57~(3), 634--655.

\bibitem[{Satake(1982)}]{Satake:1982a}
Satake, M., 1982. Fabric tensor in granular materials. In: Vermeer, P.~A.,
  Luger, H.~J. (Eds.), Proc. IUTAM Symp. on Deformation and Failure of Granular
  Materials. A.A. Balkema, Rotterdam, pp. 63--68.

\bibitem[{Schofield and Wroth(1968)}]{Schofield:1968a}
Schofield, A.~N., Wroth, P., 1968. Critical state soil mechanics. McGraw-Hill,
  New York.

\bibitem[{Shannon(1948)}]{Shannon:1948a}
Shannon, C.~E., 1948. A mathematical theory of communication. Bell System
  Technical Journal 27~(3), 379--423.

\bibitem[{Thornton(2000)}]{Thornton:2000a}
Thornton, C., 2000. Numerical simulations of deviatoric shear deformation of
  granular media. G{\'{e}}otechnique 50~(1), 43--53.

\bibitem[{Tordesillas et~al.(2011)Tordesillas, Lin, Zhang, Behringer, and
  Shi}]{Tordesillas:2011a}
Tordesillas, A., Lin, Q., Zhang, J., Behringer, R.~P., Shi, J., 2011.
  Structural stability and jamming of self-organized cluster conformations in
  dense granular materials. J. Mech. Phys. Solids 59~(2), 265--296.

\bibitem[{Troadec et~al.(2002)Troadec, Radjai, Roux, and
  Charmet}]{Troadec:2002a}
Troadec, H., Radjai, F., Roux, S., Charmet, J.~C., 2002. Model for granular
  texture with steric exclusion. Phys. Rev. E 66, 041305.

\bibitem[{Williams and Nabha(1997)}]{Williams:1997c}
Williams, J.~R., Nabha, R., 1997. Coherent vortex structures in deforming
  granular materials. Mech. Cohesive-Frict. Mater. 2~(3), 223--236.

\bibitem[{Yoon and Gim{\'{e}}nez(2012)}]{Yoon:2012a}
Yoon, S.~W., Gim{\'{e}}nez, D., 2012. Entropy characterization of soil pore
  systems derived from soil-water retention curves. Soil Sci. 177~(6),
  361--368.

\bibitem[{Zhao and Guo(2013)}]{Zhao:2013a}
Zhao, J., Guo, N., 2013. A new definition on critical state of granular media
  accounting for fabric anisotropy. In: AIP Conference Proceedings. Vol. 1542.
  pp. 229--232.

\end{thebibliography}
%
%sagemathcloud={"zoom_width":135}
%

\end{document}